%% file: RENO_LiHe.tex
\documentclass[aps,prd,twocolumn,showpacs,superscriptaddress,showkeys,bm,amsmath,amssymb]{revtex4-2}

\usepackage{graphicx} % needed for pdflatex (.pdf figures)
\usepackage{bmpsize}
\usepackage{multirow}
\usepackage{dcolumn}   % needed for some tables
\usepackage{bm}        % for math
\usepackage{amsmath}
\usepackage{amssymb}   % for math
\usepackage{amsopn}
\usepackage{upgreek}
\usepackage{here}
\usepackage{amsmath}
\usepackage{kotex}%hglee modi
\usepackage{color} %hglee modi
\usepackage{lineno}
\usepackage{hyperref}
\hypersetup{
    colorlinks=true,
    citecolor=blue,
    linkcolor=black,
    filecolor=black,      
    urlcolor=blue,
    pdftitle={RENO_LiHe},
    pdfpagemode=FullScreen,
    }

\def\Li{$^{9}\text{Li}$}
\def\He{$^{8}\text{He}$}

\def\SIM{$\sim$}

\def\CS{$^{12}$C }

\def\CfS{$^{252}$Cf }
\def\BS{$^{12}$B }
\def\NS{$^{12}$N }

\def\LiS{$^{9}\text{Li}$ }
\def\HeS{$^{8}\text{He}$ }

\def\LIS{$^{9}\text{Li}$ }

\begin{document}
\title{Measurement of cosmogenic \LiS and \HeS production rates at RENO}
\newcommand{\CNU}{\affiliation{Institute for Universe and Elementary Particles, Chonnam National University, Gwangju 61186, Korea}}
\newcommand{\DSU}{\affiliation{Institute for High Energy Physics, Dongshin University, Naju 58245, Korea}}
\newcommand{\GIST}{\affiliation{GIST College, Gwangju Institute of Science and Technology, Gwangju 61005, Korea}}
\newcommand{\IBS}{\affiliation{Institute for Basic Science, Daejeon 34047, Korea}}
\newcommand{\KAIST}{\affiliation{Department of Physics, Korea Advanced Institute of Science and Technology, Daejeon 34141, Korea}}
\newcommand{\KNU}{\affiliation{Department of Physics, Kyungpook National University, Daegu 41566, Korea}}
\newcommand{\SNU}{\affiliation{Department of Physics and Astronomy, Seoul National University, Seoul 08826, Korea}}
\newcommand{\SYU}{\affiliation{Department of Fire Safety, Seoyeong University, Gwangju 61268, Korea}}
\newcommand{\SKKU}{\affiliation{Department of Physics, Sungkyunkwan University, Suwon 16419, Korea}}

\author{H.~G.~Lee}\SKKU 
\author{J.~H.~Choi}\DSU
\author{H.~I.~Jang}\SYU
\author{J.~S.~Jang}\GIST
\author{S.~H.~Jeon}\SKKU
\author{K.~K.~Joo}\CNU
\author{D.~E.~Jung}\SKKU
\author{J.~G.~Kim}\SKKU
\author{J.~H.~Kim}\SKKU
\author{J.~Y.~Kim}\CNU
\author{S.~B.~Kim}\SKKU
\author{S.~Y.~Kim}\SNU
\author{W.~Kim}\KNU
\author{E.~Kwon}\SKKU
\author{D.~H.~Lee}\SKKU
\author{W.~J.~Lee}\SNU
\author{I.~T.~Lim}\CNU
\author{D.~H.~Moon}\CNU
\author{M.~Y.~Pac}\DSU
\author{J.~S.~Park}\KNU
\author{R.~G.~Park}\CNU
\author{H.~Seo}\SNU
\author{J.~W.~Seo}\SKKU
\author{C.~D.~Shin}\CNU
\author{B.~S.~Yang}\SNU
\author{J.~Yoo}\SNU
\author{S.~G. Yoon}\SNU
\author{I.~S.~Yeo}\DSU
\author{I.~Yu}\SKKU

\collaboration{The RENO Collaboration}
\noaffiliation
\date{\today}

%----------------------- abstract
\begin{abstract}

We report the measured production rates of unstable isotopes \LIS and \HeS produced by cosmic muon spallation on \CS using two identical detectors of the RENO experiment. Their $\beta$ decays accompanied by a neutron make a significant contribution to backgrounds of reactor antineutrino events in precise determination of the smallest neutrino mixing angle. The mean muon energy of its near\,(far) detector with an overburden of 120\,(450)\,m.w.e. is estimated as $33.1\pm2.3\,(73.6\pm4.4)$\,GeV. Based on roughly 3100 days of data, the cosmogenic production rate of \Li\,(\He) isotope is measured  to be  $44.2\pm3.1\,(10.6\pm7.4)$\,per day at near detector and $10.0\pm1.1\,(2.1\pm1.5)$\,per day at far detector. This corresponds to yields of \Li\,(\He), $4.80\pm0.36\,(1.15\pm0.81)$ and $9.9\pm1.1\,(2.1\pm1.5)$ at near and far detectors, respectively, in a unit of $10^{-8}$ 
$\mu^{-1}\rm g^{-1}\rm cm^{2}$. Combining the measured \LiS yields with other available underground measurements, an excellent power-law relationship of the yield with respect to the mean muon energy is found to have an exponent of $\alpha=0.75\pm0.05$.

\end{abstract}

\newpage
\maketitle

%------------------------Introduction
\section{Introduction}

Cosmic muons create spallation products  in underground detectors and their surrounding rocks. They produce unstable radioisotopes and neutrons by interacting with \CS in liquid scintillator detectors. Cosmogenic isotopes of \LiS and \HeS are  sources of the most serious background in the reactor neutrino oscillation measurements\,\cite{RENO:2018dro, DayaBay:2018yms, DoubleChooz:2019qbj} and potential sources of background in the double-beta decay and dark matter experiments.

The muon-induced production of radioactive isotopes was studied by an experiment\,\cite{Hagner:2000xb} at CERN using the Super Proton Synchrotron muon beam. The muon energy dependence of the spallation production was obtained for energies of 100 and 190\,GeV. The production cross sections at other energies were estimated from the measured results by extrapolation, assuming a power-law dependence on the muon energy. However, the production rate of \LiS and \HeS was measured only at 190\,GeV. The cosmogenic yield of radioactive isotopes can be calculated with a large uncertainty, using simulations based on MUSIC\,\cite{Kudryavtsev:2008qh}, FLUKA\,\cite{Ferrari:2005zk}, and GEANT4\,\cite{Allison:2006ve}.

The $\beta$ decays of \LiS and \HeS accompanied with an emitted neutron mimic an inverse beta decay\,(IBD) reaction of electron antineutrino, $\overline{\nu}_e + p \rightarrow e^{+} + n$. Their relatively long lifetime and high muon rate of shallow detector sites make it difficult to separate the \LiS and \HeS decays from the IBD events using a timing veto criterion. Therefore, accurate production rates of the cosmogenic isotopes are of great interest to make a precise determination of the neutrino mixing angle $\theta_{13}$ as well as to make a sensitive search for neutrinoless-double-beta decay and dark matter. 
   
The cosmogenic \LiS and \HeS production rates have been measured and reported by the underground experiments of Borexino\,\cite{Borexino:2013cke}, Daya Bay\,\cite{DayaBay:2016ggj}, Double Chooz\,\cite{DoubleChooz:2018kvj} and KamLAND\,\cite{KamLAND:2009zwo}. The presented rates were indirectly measured from observed cosmogenic neutron and \BS production, or obtained by the delayed time with respect to a parent muon. However, correct association between a cosmogenic isotope and its preceding muon is not possible when it comes to a high muon rate relative to the \LiS or \HeS lifetimes. This paper presents a direct measurement of the cosmogenic \LiS and \HeS production rates in two identical RENO detectors with 120 and 450\,m.w.e. overburdens. We employ a method of measuring the  rates from the observed $\beta$-decay spectra of the isotopes produced by the entire cosmic muons passing through the detectors. 
   
Section II presents an overview of the RENO experiment. Section III describes detection of cosmic muons and produced cosmogenic radioisotopes. Section IV describes selection criteria for the cosmogenic \LiS and \HeS $\beta$ decays and their backgrounds. Section V presents their measured $\beta$-decay spectra. In Sec. VI, we report the observed and production rates of cosmogenic \LiS and \HeS isotopes in the two RENO detectors. Finally, we summarize in Sec. VII.

%----------------------The RENO experiment
\section{The RENO experiment}

The RENO experiment measured the smallest neutrino mixing angle $\rm{\theta_{13}}$ based on the disappearance of electron antineutrinos produced by the Hanbit nuclear reactors\,\cite{RENO:2012mkc,RENO:2018dro}. Two identical near and far detectors\,(ND and FD) were deployed at distances of 290 and 1380\,m from the reactor array center, respectively. They detect the reactor antineutrinos through the IBD reaction,  using liquid scintillator\,(LS) loaded with 0.1\% gadolinium\,(Gd) as a target. The interaction is identified by coincidence between the prompt positron signal and the delayed neutron capture on Gd. 

\begin{figure}[h]
\begin{center}
\includegraphics[width=0.48\textwidth]{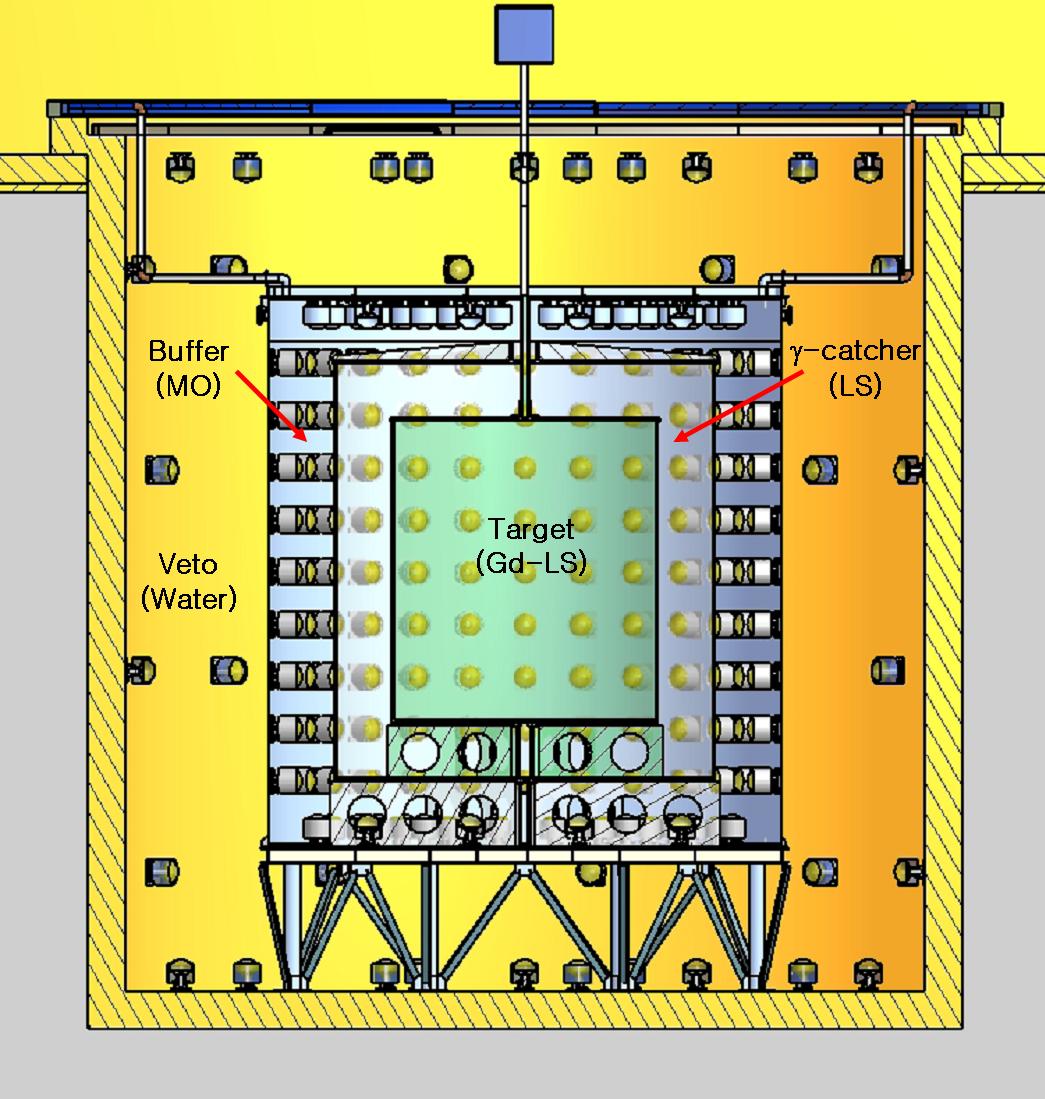}% RENO detector
\end{center}
\caption{Schematic view of the RENO detector. A main inner detector\,(ID) consists of the neutrino target, $\gamma$ catcher and buffer from the innermost and is contained in a cylindrical stainless steel vessel. An outer detector\,(OD) of veto surrounds the ID and is filled with highly purified water. \label{figure1}
}\end{figure}

The detector consists of four cylindrical layers as shown in Fig.\,\ref{figure1}. They are neutrino target, $\gamma$ catcher, buffer and veto from the innermost and filled with different liquids. A main inner detector\,(ID) consisting of the neutrino target, $\gamma$ catcher and buffer, is contained in a cylindrical stainless steel vessel of 5.4\,m in diameter and 5.8\,m in height and houses two nested cylindrical acrylic vessels. The innermost target vessel, a 25\,mm thick acrylic vessel of 2.75\,m in diameter and 3.15\,m in height, holds 16\,ton of 0.1\% Gd-doped LS\,(Gd-LS) as a neutrino target. It is surrounded by a 60\,cm thick layer of 29\,ton undoped LS in $\gamma$ catcher, useful for recovering $\gamma$ rays escaping from the target region. The $\gamma$-catcher liquid is contained in a 30\,mm thick acrylic vessel of 4.0\,m in diameter and 4.4\,m in height. Outside the $\gamma$ catcher is a 70\,cm thick buffer region filled with 65\,ton of mineral oil\,(MO). It provides shielding against ambient $\gamma$ rays and neutrons coming from outside. An outer detector\,(OD) of 1.5\,m in thickness surrounds the ID and is filled with 350 ton of highly purified water.
 
 Light signals emitted from particles interacting in ID are detected by a total of 354 low-background 10-inch Hamamatsu R7081 photomultiplier tubes\,(PMTs)\,\cite{Ma:2009aw} that are mounted on the inner wall of the ID. The OD is equipped with 67 10-inch R7081 water-proof PMTs mounted on the wall of the concrete veto vessel. The inner surface of OD is covered with Tyvek sheets to increase the light collection.

\vspace{4mm}
%-------------------------Cosmic muons and cosmogenic radioisotopes
\section{Cosmic muons and cosmogenic radioisotopes}

%-------------------------
\subsection{Cosmic muon rate}

\begin{figure}
\begin{center}
\includegraphics[width=0.5\textwidth]{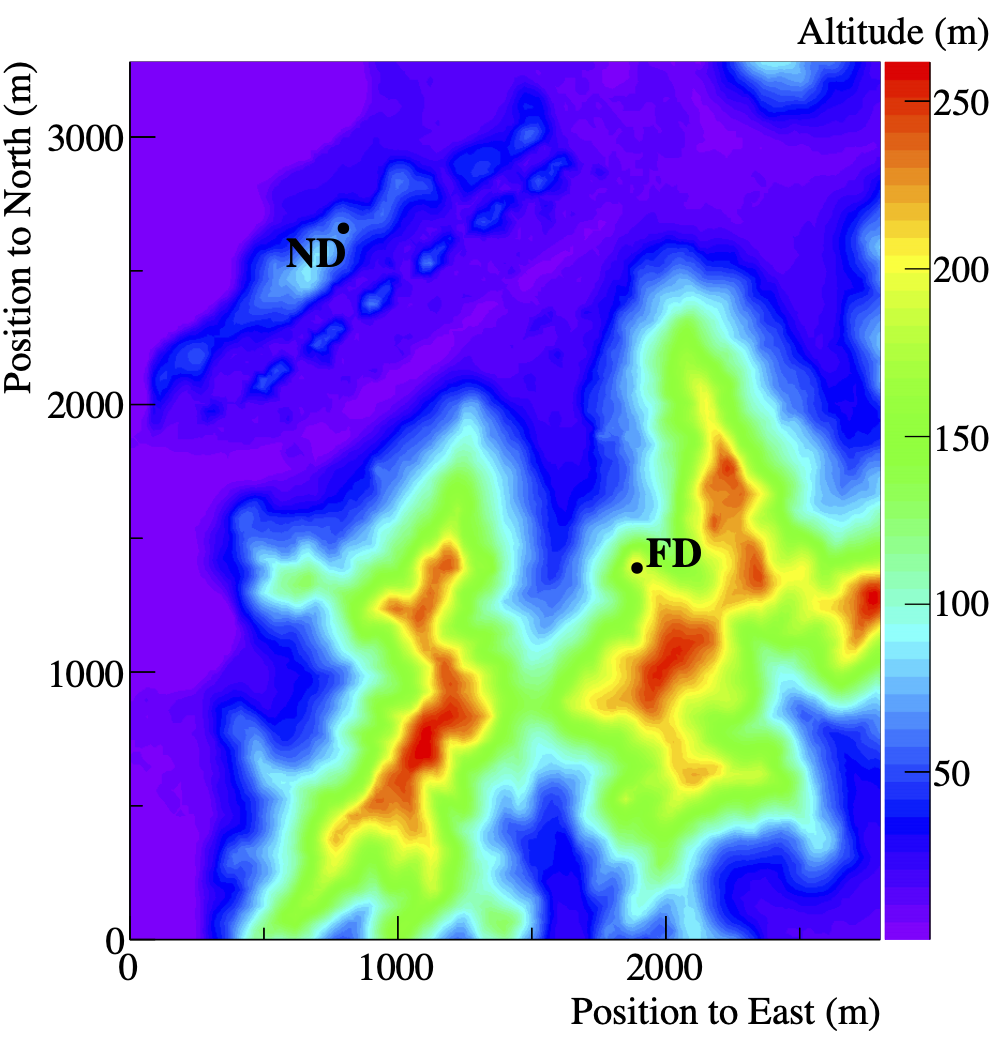}% local mountain distribution
\end{center}
\caption{Topological profile of the RENO detector site. The black points indicate locations of the RENO near and far detectors. The origin of coordinates corresponds to $126.41^\circ$ in longitude and $35.39^\circ$ in latitude.}  
\label{figure2}
\end{figure}

\begin{figure}[h]
\begin{center}
\includegraphics[width=0.48\textwidth]{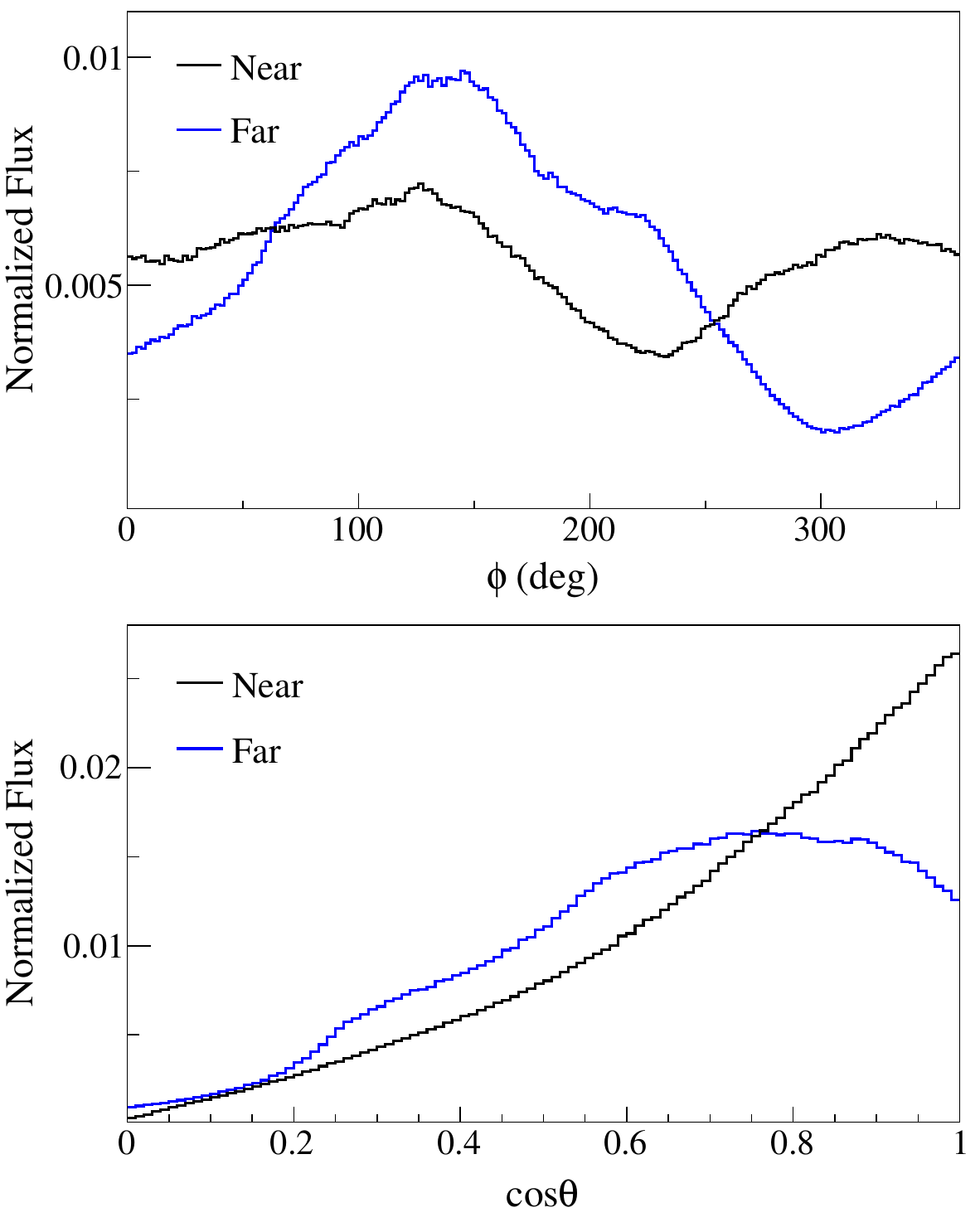}% phi cos distribution
\end{center}
\caption{Expected angular distributions of muons passing through ND and FD as a function of azimuthal\,($\mathrm \phi$) and zenith angles\,$(\mathrm \theta)$. The angle $\phi$ is defined with respect to the east direction. }
\label{figure3}
\end{figure}

 Both underground ND and FD observe cosmic muon fluxes attenuated by their overburdens. According to our Monte Carlo simulation\,(MC) using the MUSIC package\,\cite{Kudryavtsev:2008qh}, the mean muon energies\,($\overline{E}_{\rm \mu}$) are $33.1\pm2.3$ and $73.6\pm4.4$\,GeV for the ND and FD sites, respectively, whereas the energy at ground is roughly 1\,GeV.  The mean muon energy increases as function of overburden. The modified Gaisser parametrization\,\cite{Tang:2006uu} is used as an initial sea-level muon flux. The local mountain profile is obtained using a topographic map of ALOS\,World\,3D\,\cite{tadono2016generation} and shown in Fig.\,\ref{figure2}. An average rock density of $\rm 2.74 g/cm^3$, obtained from a geological survey, and a standard rock composition are used for muon transportation in the slant depth of the RENO site. The error of the mean muon energy is estimated by uncertainties associated with mountain profile, rock density, detector location and simulated map range, as listed in Table\,\ref{t:muon_E_err}. The expected angular distributions of cosmic muons passing through ND and FD are obtained by the MC and shown in Fig.\,\ref{figure3} where they reflect the influence of the local geographical topology.

\begin{table}[h]
\begin{center}
\caption{\label{t:muon_E_err} 
 Uncertainties of mean muon energy.}
\begin{tabular*}{0.48\textwidth}{@{\extracolsep{\fill}} lcc}
\hline
\hline
Uncertainty source & ND\,(\%) & FD\,(\%)  \\

\hline
Mountain profile    			& 5.3 & 4.1  \\
Rock density  				& 3.7 & 3.5  \\
Detector location   			& 1.7 & 1.5  \\
Simulated map range    		& 2.0 & 2.0  \\
\hline
Total uncertainty   			& 7.0 & 5.9 \\
\hline
\hline
\end{tabular*}
\end{center}
\end{table}

Cosmic muons are identified by the Cherenkov light in the OD water; the scintillation and Cherenkov light in the ID organic liquids. The muon visible energy\,$(E_{\rm\mu}^{\rm vis})$ is observed by scintillating light above 60\,MeV. On the other hand, it is observed predominantly by Cherenkov light below 60\,MeV.

\begin{figure}
\includegraphics[width=0.485\textwidth]{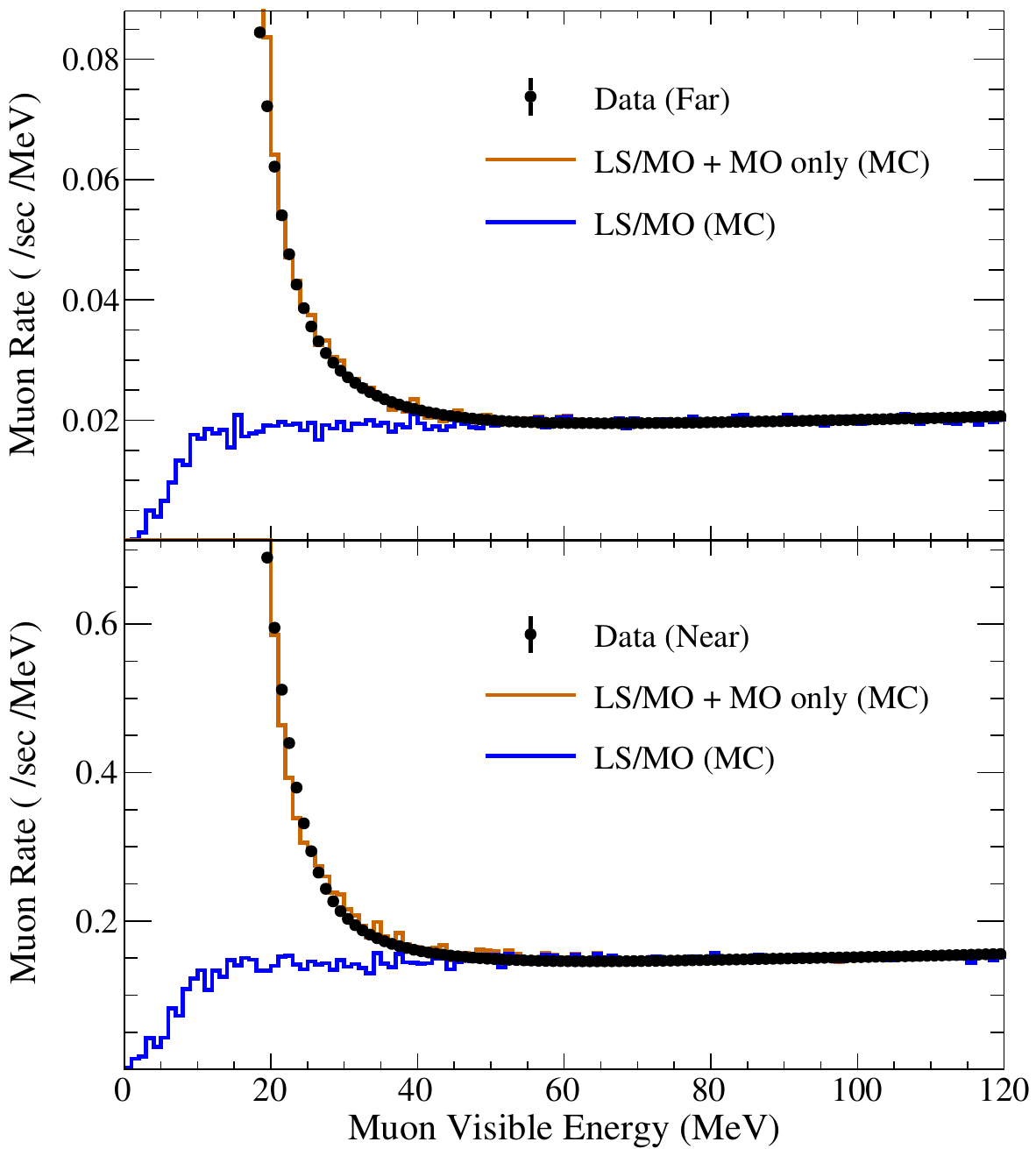}% muon rate estimation
\caption{Visible energy distributions of cosmic muons observed in FD and ND.} 
\label{figure4}
\end{figure}

Almost all of cosmic muons are well identified by requiring the number of hit OD PMTs\,(NHIT) greater than or equal to 15. Figure\,\ref{figure4} shows the observed  $E_{\rm\mu}^{\rm vis}$ distribution of cosmic muons with NHIT\,$\geq$\,15 in FD. The cosmic muons passing through the buffer region generate only Cherenkov light in the mineral oil, and thus dominate the cosmic muon rate below 40\,MeV. The rate of cosmic muons traversing the target or $\gamma$-catcher regions below 70\,MeV is estimated by extrapolation of a fit to the data above 70 MeV using a MC predicted energy distribution. The observed rate of cosmic muons traversing the $\gamma$-catcher region is $125.7\pm1.4$\,$\rm s^{-1}$ at ND and $14.0\pm0.2$\,$\rm s^{-1}$ at FD, corresponding to fluxes of $6.67\pm0.15$ and  $0.74\pm0.02\,\rm m^{-2}s^{-1}$, respectively. Using a MC calculation, the rate of cosmic muons traversing both target and $\gamma$-catcher regions is estimated to be $61.8\pm0.7$\,$\rm s^{-1}$ at ND and $6.9\pm0.1$\,$\rm s^{-1}$ at FD. 

\begin{figure}[h]
\includegraphics[width=0.485\textwidth]{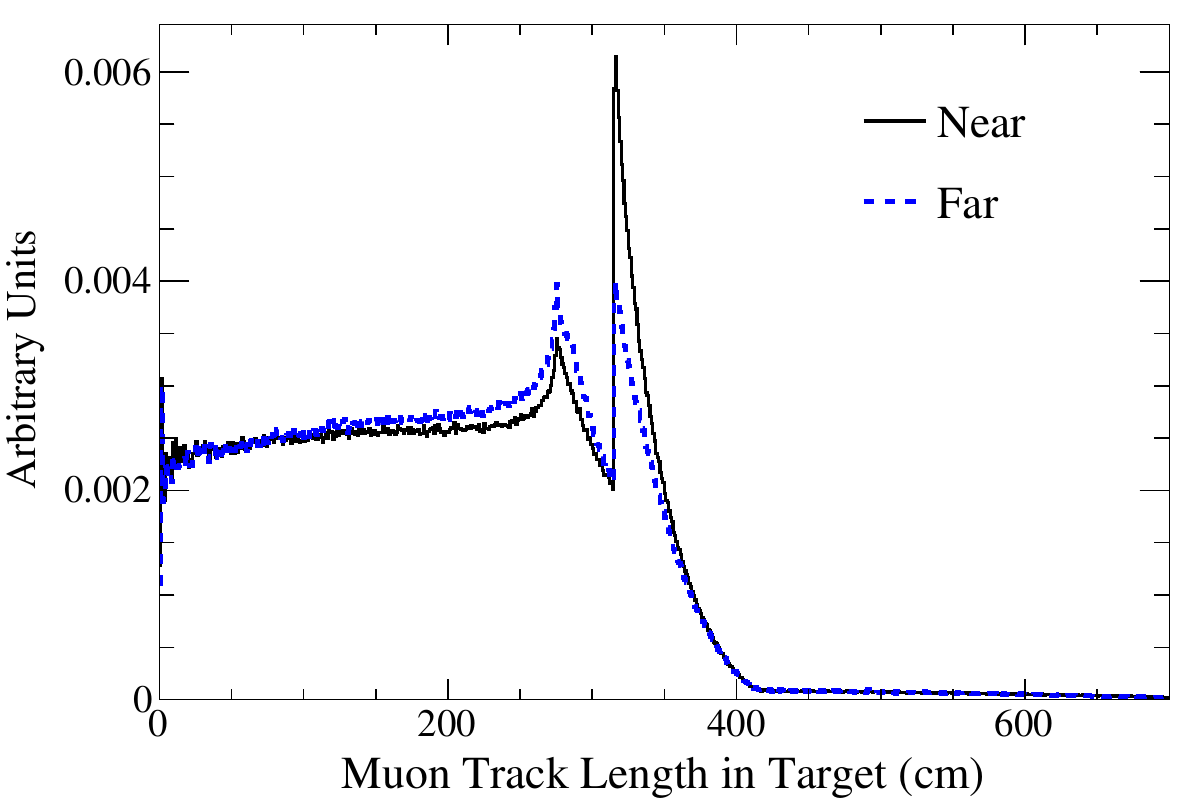} %Average muon track length
\caption{ Distributions of muon track lengths in the target region. The track lengths are obtained from the MC simulation. A peak at 275\,(315)\,cm comes from muons passing 
%in the barrel-to-barrel\,(top-to-bottom) direction 
in the barrel-to-barrel\,(top-to-bottom), i.e.,  lateral\,(vertical) direction 
of the cylindrical target.}
\label{figure5}
\end{figure}

An average muon track length contained in the target region is estimated by a toy MC simulation and obtained as $201.1\pm3.9$\,cm at ND and $197.4\pm3.9$\,cm at FD according to the calculated track distributions as shown in Fig.\,\ref{figure5}. The error of the average muon track length is due to uncertainties of detector dimension, rock density profile, mountain profile, topographic map and its range, detector position, \rm{d}\it{E}/\rm{d}\it{X}\rm{}, and muon multiplicity as listed in Table\,\ref{t:track_length}. 

\begin{table}[h!]
\caption{ \label{t:track_length} Uncertainties of average muon track length. }
\begin{center}
\begin{tabular*}{0.48\textwidth}{@{\extracolsep{\fill}} lc}
\hline
\hline
Uncertainty source & Fractional error\,(\%)   \\
\hline
MC statistics   & 0.15   \\
Detector dimension   & 0.12   \\
Rock density profile  & 0.16  \\
Mountain profile   & 0.25  \\
Topographic map range   & 0.01   \\
Detector position   & 0.21   \\
d\it{E}/\rm{d}\it{X}\rm{} & 0.24   \\
Muon multiplicity & 1.90   \\
\hline
Total uncertainty & 1.96 \\
\hline
\hline
\end{tabular*}
\end{center}
\end{table}

\vspace{5mm}

%-------------------------
\subsection{Cosmogenic radioisotopes}

 The KamLAND and Borexino underground detectors observed various radioactive isotopes, including $\rm ^{12}B$, $\rm ^{12}N$, \LiS and \He, as spallation products of high-energy cosmic muons\,\cite{KamLAND:2009zwo, Borexino:2013cke}.  A dominant production mechanism of the cosmogenic radioisotopes is understood as the fragmentation of \CS by muon-induced hadronic showers, mostly by $\pi^{-}$\,\cite{Hagner:2000xb}. Both RENO ND and FD detectors have also observed $\beta^-$ decays of $\rm ^{12}B$\,($\tau=29.1$\,ms, $\textit{Q}=13.4$\,MeV) and $\beta^+$ decays of $\rm ^{12}N$\,($\tau=15.9$\,ms, $\textit{Q}=17.3$\,MeV) as a result of cosmogenic production\,\cite{Ajzenberg-Selove:1990fsm}. Their decay candidates are selected by requiring an energy of $3<E<20$\,MeV and an elapsed time of $\Delta T < 500$\,ms from its preceding muon which has a visible energy larger than 1.6\,(1.5)\,GeV at ND\,(FD). 
   
\begin{figure}[h]
\includegraphics[width=0.495\textwidth]{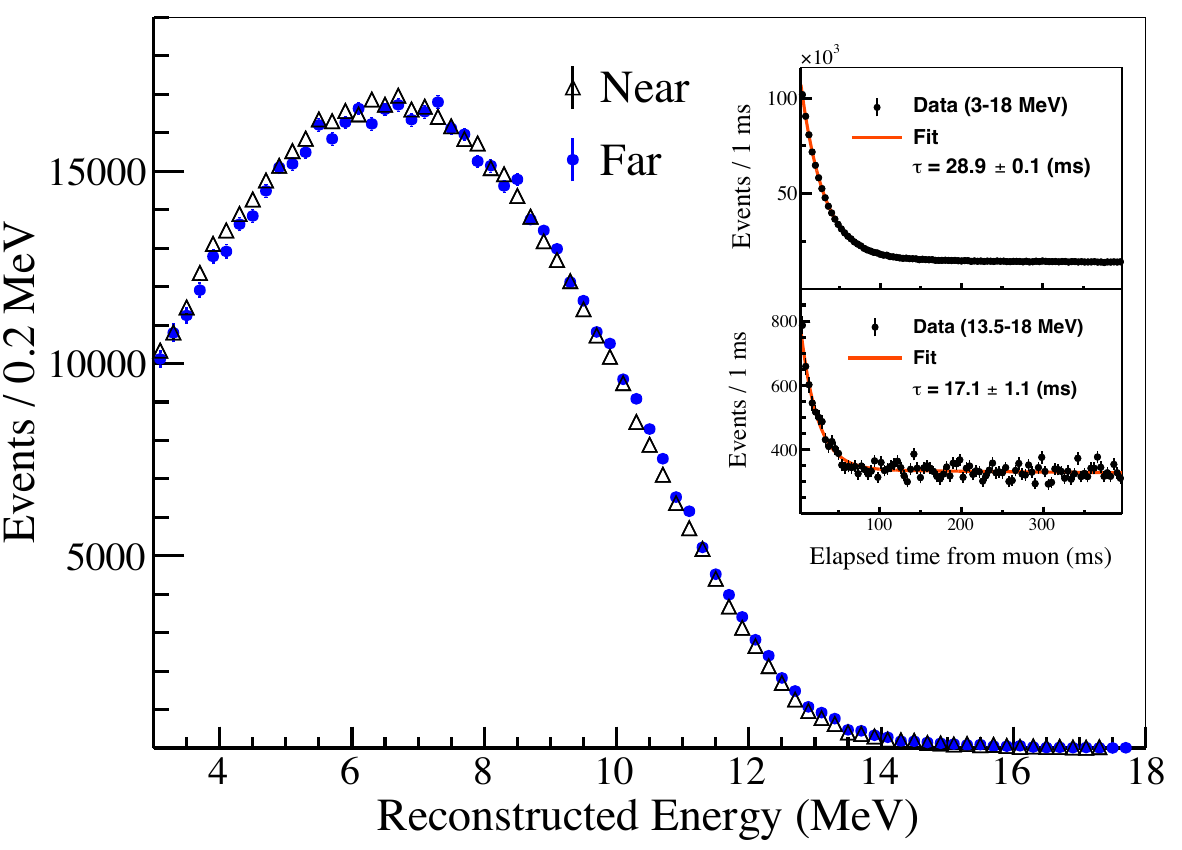} 
\caption{Observed energy spectra and $\Delta T$ distributions\,(inset) of $\beta$ decays from cosmogenic $\rm ^{12}B$ and $\rm ^{12}N$ isotopes.}
\label{figure6}
\end{figure}
\vspace{0.65mm}

Figure\,\ref{figure6} shows observed energy spectra and $\Delta T$ distributions of $\beta$ decays from the cosmogenic \BS and \NS isotopes in the RENO detectors. The observed spectra are obtained from the selected events with $2 < \rm{\Delta}  \textit{T} < 60$\,ms.  The longer lived isotopes of \He, \Li, $\rm ^{9}C$, $\rm ^{8}Li$ and $\rm ^{8}B$ also contribute to the selected sample with roughly constant decay rates on this timescale. Their contribution is subtracted from  the spectra using a fit-out as a longer decay-time component. The elapsed time  distribution of $\Delta T$ is distorted if the muon rate surpasses the decay rates of $\rm ^{12}B$ and $\rm ^{12}N$.  The muon rate of $E_{\mu}^{\rm vis}>1$\,GeV is $1.26\pm0.03\,(0.29\pm0.02)$ $\rm s^{-1}$ at ND\,(FD) and does not result in the $\Delta T$ distortion. The observed $\beta$-decay rate of unstable $\rm^{12}B$ plus $\rm^{12}N$ isotopes, produced by muons with $E_{\mu}^{\rm vis}>1$\,GeV, is measured to be $481.1\pm3.1\,(179.2\pm1.5)$\,per day in the ND\,(FD) target and  $\rm\gamma$-catcher region. 

 Neutron-unstable excited states of the cosmogenic \LiS and \HeS isotopes are also produced in the RENO detectors as a serious background against detecting the IBD events of reactor antineutrinos. They decay into daughter nuclei with neutron-unstable excited states by emitting a $\beta^-$\,particle\,($\tau=257.2$\,ms, $\textit{Q}=13.6$\,MeV for \LiS or $\tau=171.7$\,ms, $\textit{Q}=10.7$\,MeV for \He) as well as a neutron from the subsequent decay\,\cite{Tilley:2004zz}. The neutron is captured mainly on Gd with a mean capture time of \SIM\,26\,$\mu$s in the target region\,\cite{RENO:2016ujo}.  The coincidence of a $\beta$ emission and a delayed neutron capture provides a clean signature of cosmogenic \LiS and \HeS production in the RENO detector.

%-------------------------Events selection
\section{Events selection }

%-------------------------
\subsection{Data sample }
 In this analysis $\sim$3100 live days of data taken from August 2011 to August 2020 are used for measuring the cosmogenic \LiS and \HeS yields at RENO. We require a prompt signal from a $\beta$ decay and a delayed signal from a neutron capture by Gd in the entire target region in order to obtain $\beta$-n emitters from \LiS and \He. The selection criteria for the sample are identical to those for the reactor antineutrino events, except for releasing a muon time veto requirement that is applied for removing cosmogenic spallation products. Event reconstruction and energy calibration are described in detail elsewhere\,\cite{RENO:2016ujo}.

%-------------------------
\subsection{Background }
 There are correlated and uncorrelated backgrounds between the prompt and delayed candidates. The correlated background comes from reactor electron  antineutrinos and fast neutrons. The reactor antineutrino IBD events cannot be discriminated against the \LiS and \HeS $\beta$-n emitters because of their identical prompt and delayed signals. However, the IBD prompt spectrum is rapidly depleted above 8\,MeV  while the \Li\,(\He) $\beta$ spectrum extends up to 14\,(11)\,MeV.  The fast neutrons are produced by cosmic muons traversing the detector and the surrounding rock. An energetic neutron entering the ID interacts in the LS to produce a recoil proton before being captured. The recoil proton generates scintillation light mimicking a promptlike event. The uncorrelated background is due to random association between the prompt and delayedlike candidates, also called ``accidental background''. The accidental promptlike events come mostly from ambient $\gamma$ rays with energies less than $\sim3$\,MeV. A detailed description of IBD backgrounds is given in Ref.\,\cite{RENO:2016ujo}. 
 
In contrast to the IBD sample of Ref.\,\cite{RENO:2016ujo}, this analysis requires minimum muon veto criteria to collect the $\beta$-n emitters from cosmogenic 
\LiS and \He. Therefore, either promptlike or delayedlike events can come from $\beta$ decays of various cosmogenic radioisotopes including $\rm ^{12}B$ with the highest production rate. Such a delayed event may be accidentally paired with an ambient $\gamma$ ray as a prompt event to form an uncorrelated background. A high-energy muon occasionally produces multiple $\rm ^{12}B$'s or a $\rm ^{12}B$ plus a longer-lived isotope\,($\rm X$), such as $\rm ^{9}C$, $\rm ^{8}Li$ and $\rm ^{8}B$, to mimic a $\beta$-n pair from the \LiS or \HeS decays. The background is called ``$\rm ^{12}B$-$\rm ^{12}B/X$ background.''  

%-------------------------
\subsection{Event selection criteria}
   The following criteria are applied to obtain a sample of $\beta$-n emitters from cosmogenic \LiS and \HeS isotopes: (i) a prompt energy ($E_{\rm p}$) requirement of $1.2 < E_{\rm p} < 15$\,MeV; (ii) a delayed energy ($E_{\rm d}$) requirement of $6 < E_{\rm d} < 12$\,MeV; (iii) a time coincidence requirement of $2 < \Delta T_{\rm pd} < 100\,\mu s$, where $\Delta T_{\rm pd}$ is the time difference between prompt and delayed candidates; (iv) a spatial coincidence requirement of $\Delta R < 2$\,m where $\Delta R$  is the distance between prompt and delayed candidates. Most of other selection criteria for the reactor antineutrino events are applied and given in Ref.\,\cite{RENO:2016ujo}. The timing veto criteria of rejecting events associated with cosmic muons, are not applied in order to keep the \LiS and \HeS $\beta$-n decays. However, a timing veto requirement is imposed to reject immediate shower events if they are within a 1\,ms window following a cosmic muon of visible energies larger than 70\,MeV or of energies between 20 and 70\,MeV for OD PMT hits greater than 50. After applying all the selection criteria, we obtain 1\,211\,335\,(158\,938) $\beta$-n candidates at ND\,(FD) detector from 3100 live days of data.

%\newpage
%----------------------------Measurement of \LiS plus \HeS spectrum
\section{Measurement of \LiS plus \HeS spectrum}

%------------------------------------
\subsection {Predicted spectra}
\begin{figure}
\includegraphics[width=0.49\textwidth]{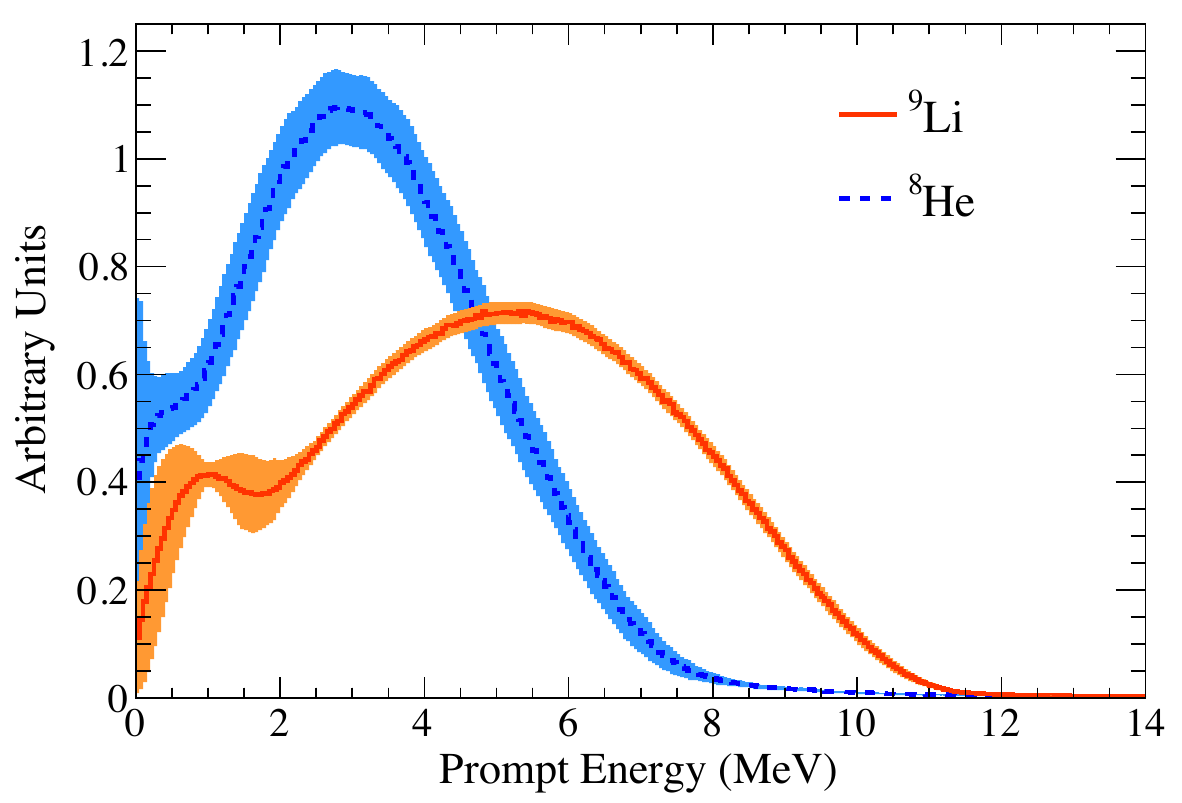} 
\caption{Predicted $\rm \beta$-decay spectra of \LiS and \HeS isotopes. Shaded bands indicate the spectral errors. The two spectra are normalized to have equal areas.}
\label{figure7}
\end{figure}

The \Li\,(\He) $\rm \beta$-decay produces neutron-unstable excited states of $\rm ^{9}Be$\,($\rm ^{8}Li$) with a branching ratio of $50.8\pm0.9\%\,(16\pm1\%)$\,\cite{Tilley:2004zz}. The daughter nuclei subsequently emit a single neutron that is thermalized and captured on Gd to release a few $\rm \gamma$ rays of $\sim$8 MeV in total energy in the target region. The $\rm \beta$ decays with an emitted neutron are accompanied by additional $\alpha$, $\gamma$, or tritium particles in the subsequent daughter-nuclei decays. A complete list of the $\rm \beta$-n decays, including branching ratios, mean values of energy levels, and decay widths of daughter nuclear states, is obtained from Ref.\,\cite{Tilley:2004zz} and used in our MC simulation. The $\rm \beta$-decay spectra are computed according to the Fermi theory with Breit-Wigner correction. The predicted spectra from the MC simulation are shown in Fig.\,\ref{figure7}, where the spectral errors are estimated from the uncertainties associated with branching ratios, weak magnetism corrections, energy scale and resolution, and the quenching effect of accompanied particles in the liquid scintillator. They are used for comparison with data as well as to extract the fractional ratios of \LiS and \HeS $\rm \beta$ decays. 

%------------------------------------
\subsection {Measured spectrum of \Li+\HeS $\beta$ decays }
\begin{figure}[h!]
\includegraphics[width=0.485\textwidth]{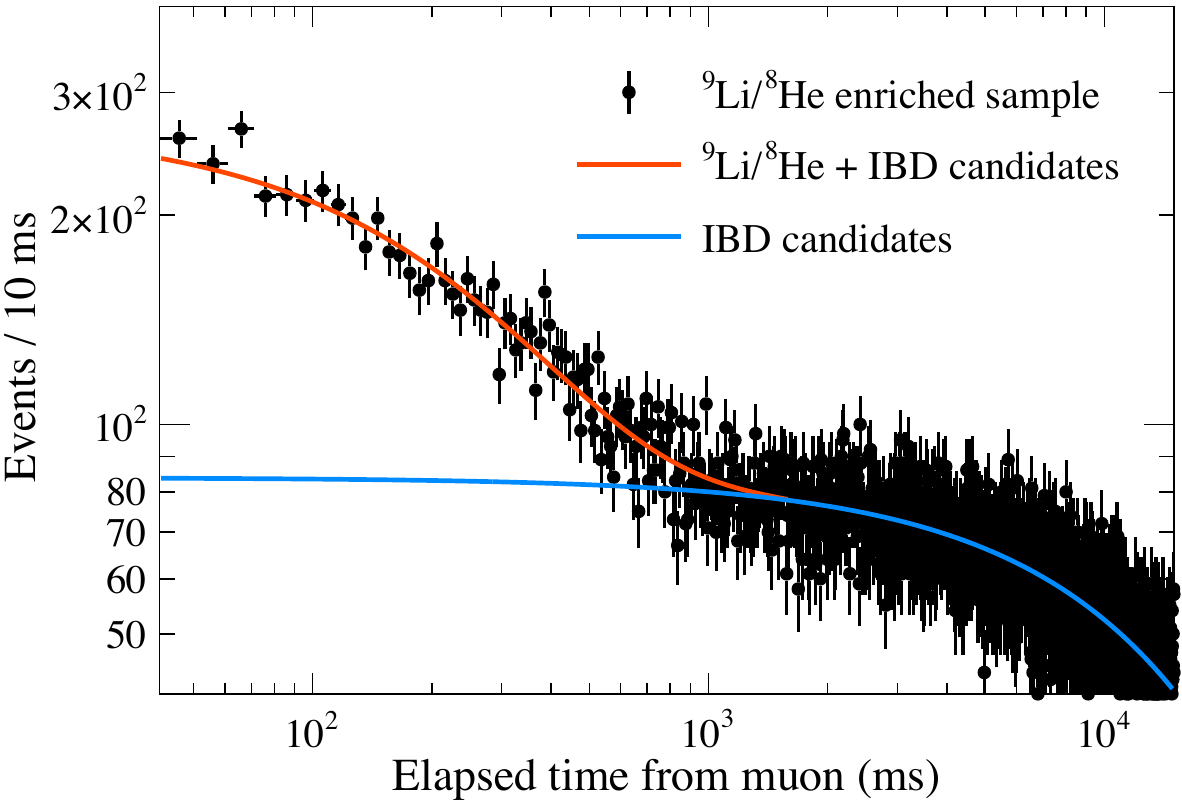} 
\caption{Elapsed time distribution of selected $\rm \beta$-n candidates in the far detector, with respect to a preceding muon with $E_{\mu}^{\rm vis}> 1.5$\,GeV, in the far detector.}
\label{figure8}
\end{figure}

A high purity of \LiS and \HeS decay sample is obtained by requiring a $\rm \beta$-n candidate to have an elapsed time of  $40 < \Delta T < 500\,(400)$\,ms from a preceding muon with $E_{\mu}^{\rm vis}> 1.5\,(1.6)$\,GeV, in FD\,(ND). Figure\,\ref{figure8} shows the elapsed-time distribution of the $\rm \beta$-n candidates before applying the requirement. The $\rm \beta$-n candidates of $\Delta T < 40$\,ms consist of $\rm ^{12}B$-$\rm ^{12}B$/X, accidental and IBD backgrounds and thus are excluded for measuring the \LiS and \HeS $\rm \beta$-decay spectrum. The accidental background comes from random coincidence between an ambient $\rm \gamma$ ray as a prompt candidate and a $\rm ^{12}B$ decay as a delayed candidate. 

\begin{figure}[h!]
\includegraphics[width=0.477\textwidth]{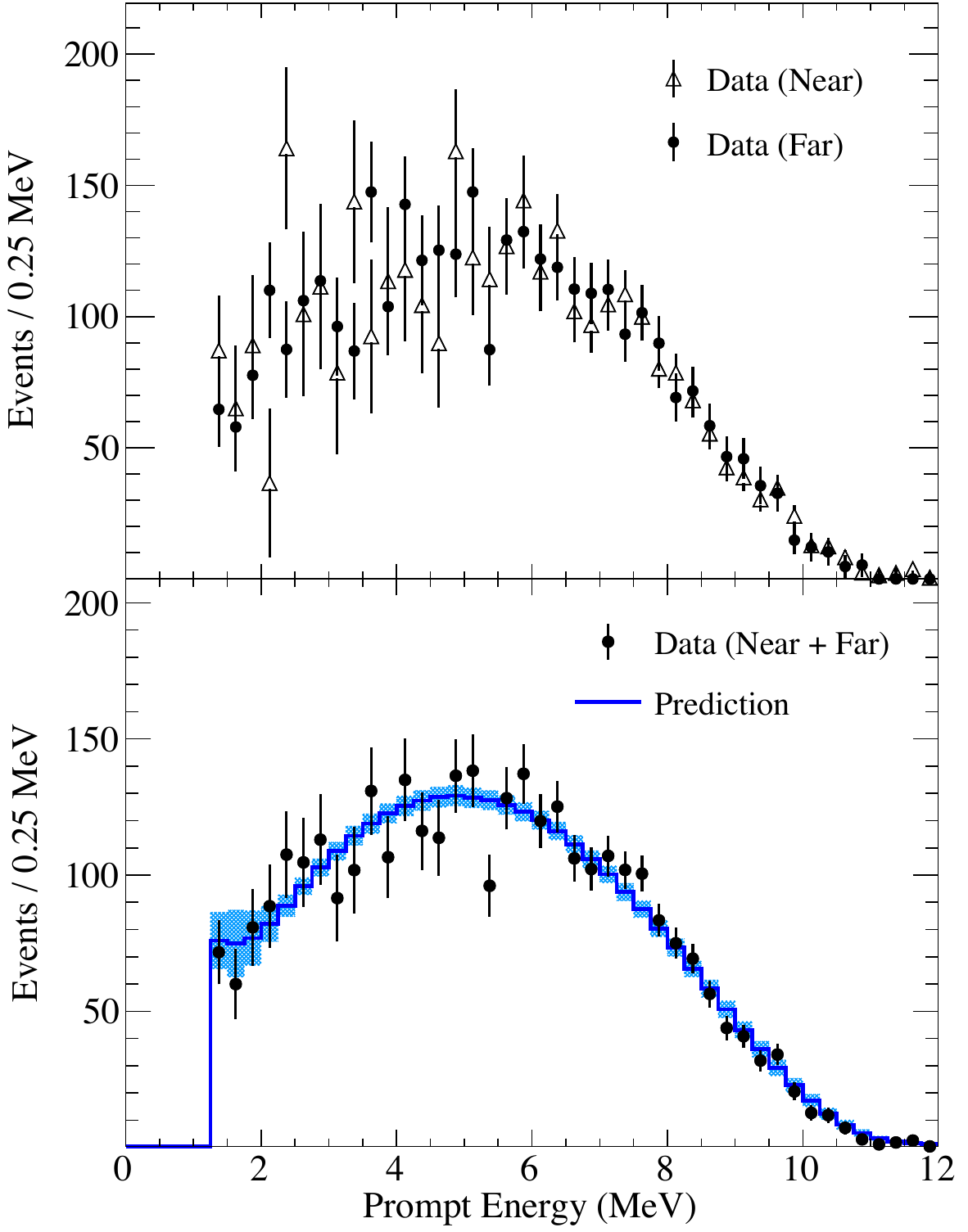} 
\caption{(Upper) Measured \Li+\HeS $\beta$-decay spectra at both detectors. (Lower) Combined spectrum in comparison with the MC prediction. The shaded band represents the shape uncertainty.}
\label{figure9}
\end{figure}

 A long-time$\,(\Delta T > 1 \, \rm s)$ component in the elapsed time distribution is dominated by reactor antineutrino IBD candidates which are uncorrelated with muons in time. The mean decay time of \Li+\HeS is measured to be $260\pm12$\,ms at FD and $258\pm20$\,ms at ND indicating predominant production of \LiS over \He. The \Li+\HeS $\rm \beta$-decay spectrum is obtained by subtracting the long-time component from the \Li+\HeS enriched sample. The magnitude of the long-time component is determined by a fit to the decay time distribution. The measured \Li+\HeS $\rm \beta$-decay spectra observed at ND and FD are shown in Fig.\,\ref{figure9}. The spectral errors come from the statistical uncertainties in the subtraction. The large spectral uncertainty below 8\,MeV is caused by subtracting the long-time spectrum of reactor antineutrino candidates, and thus the error below 8\,MeV is larger for the ND detector. On the other hand, the spectral error above 8\,MeV is smaller for the ND detector due to higher \Li+\HeS production at shallower underground. Identical \Li+\HeS $\beta$-decay spectra at both detectors are expected because of predominant production of \LiS and identically observed mean decay times. A combined spectrum of \Li+\HeS $\beta$-decays is obtained as a weighted mean of ND and FD spectra and used for this measurement. A good agreement between the combined spectrum and the MC prediction is observed as shown in Fig.\,\ref{figure9}.
      
%\newpage
%-------------------------Measurement of \LiS and \HeS production rate
\section{Measurement of \LiS and \HeS production rate}

%---------------------------------------------
\subsection {Estimated background rates }

The obtained final sample of $\beta$-n candidates contains remained backgrounds of IBD reactor neutrino events up to $E_{\rm p}\sim8$\,MeV, accidental pairs below $E_{\rm p}=3$\,MeV, and fast neutrons. Most of the backgrounds are removed by a requirement of the prompt energy larger than 8\,MeV. The total number of $\beta$-n candidate events at $8 < E_{\rm p} <15$\,MeV are 10\,608 in the ND and 5668 in the FD. The remained background rates are given in Table \ref{t:bkg}. The fast neutron rate is estimated by extrapolating from the background dominant region of 15 $<$ $E_{\rm p}$ $<$\,50\,MeV based on its flat energy spectral shape. The remained IBD candidate rate is estimated by the Huber-Mueller\,\cite{Huber:2011wv,Mueller:2011nm} predicted spectrum. A tiny $\rm ^{252}Cf$ contamination was accidentally introduced into both detectors during the radioactive source calibration in October 2012\,\cite{RENO:2016ujo}. The source container was not tightly sealed because of a loose O ring.  Most of $\rm ^{252}Cf$ contamination background events are eliminated by a multiple-neutron requirement. The remaining background rate and spectral shape are given in Ref.\,\cite{RENO:2016ujo}. In this subsection, we present estimation of the remained $\rm ^{12}B$-$\rm ^{12}B$/X background rate.

\begin{table}[ht!]
\caption{\label{t:bkg} Observed \Li+\HeS $\beta$-decay rates and estimated background rates at $8< E_{\rm{p}} <15$\,MeV. Rates are given per day. }
\begin{center}
\begin{tabular*}{0.48\textwidth}{@{\extracolsep{\fill}} lcc}
\hline
\hline
Detector & ND & FD  \\
\hline
Live time\,(days)  & 2675.50 &  3075.84 \\
%\hline
Observed $\beta$-n candidate rate  & 3.96$\pm$0.04&  1.84$\pm$0.02  \\
\hline
\Li+\HeS rate     & 1.46$\pm$0.07 &  0.37$\pm$0.04  \\
Total background rate & 2.49$\pm$0.06 & 1.46$\pm$0.09 \\
\hline
IBD   & 0.15$\pm$0.02  &  0.02$\pm$0.01  \\
Fast neutron     & 2.10$\pm$0.03 &  0.43$\pm$0.01 \\
$\rm ^{12}B$-$\rm ^{12}B/X$  & 0.05$\pm$0.01 &  0.01$\pm$0.01 \\
\CfS contamination   & 0.19$\pm$0.04 &  1.01$\pm$0.09 \\
%\hline
\hline
\hline
\end{tabular*}
\end{center}
\end{table}

  As described earlier, a $\beta$ decay from a cosmogenic radioisotope can mimic a delayed candidate of 8\,MeV $\gamma$ rays from neutron capture on Gd. A cosmic-ray muon predominantly produces $\rm ^{12}B$ as well as  $\rm ^{12}N$,  $\rm ^{9}C$, \Li, \He, $\rm ^{8}Li$ and $\rm ^{8}B$ with decay times of 16\,ms to $\sim$1\,s. All of these radioisotopes release a $\beta$ radiation between 6 and 12\,MeV, a delayed energy range, because of their end point energies larger than 10\,MeV. Such a delayed candidate forms a background pair in the \LiS and \HeS $\beta$-n final sample, as either accidental or $\rm ^{12}B$-$\rm ^{12}B$/X backgrounds. 
  
\begin{figure}[!h]
\includegraphics[width=0.48\textwidth]{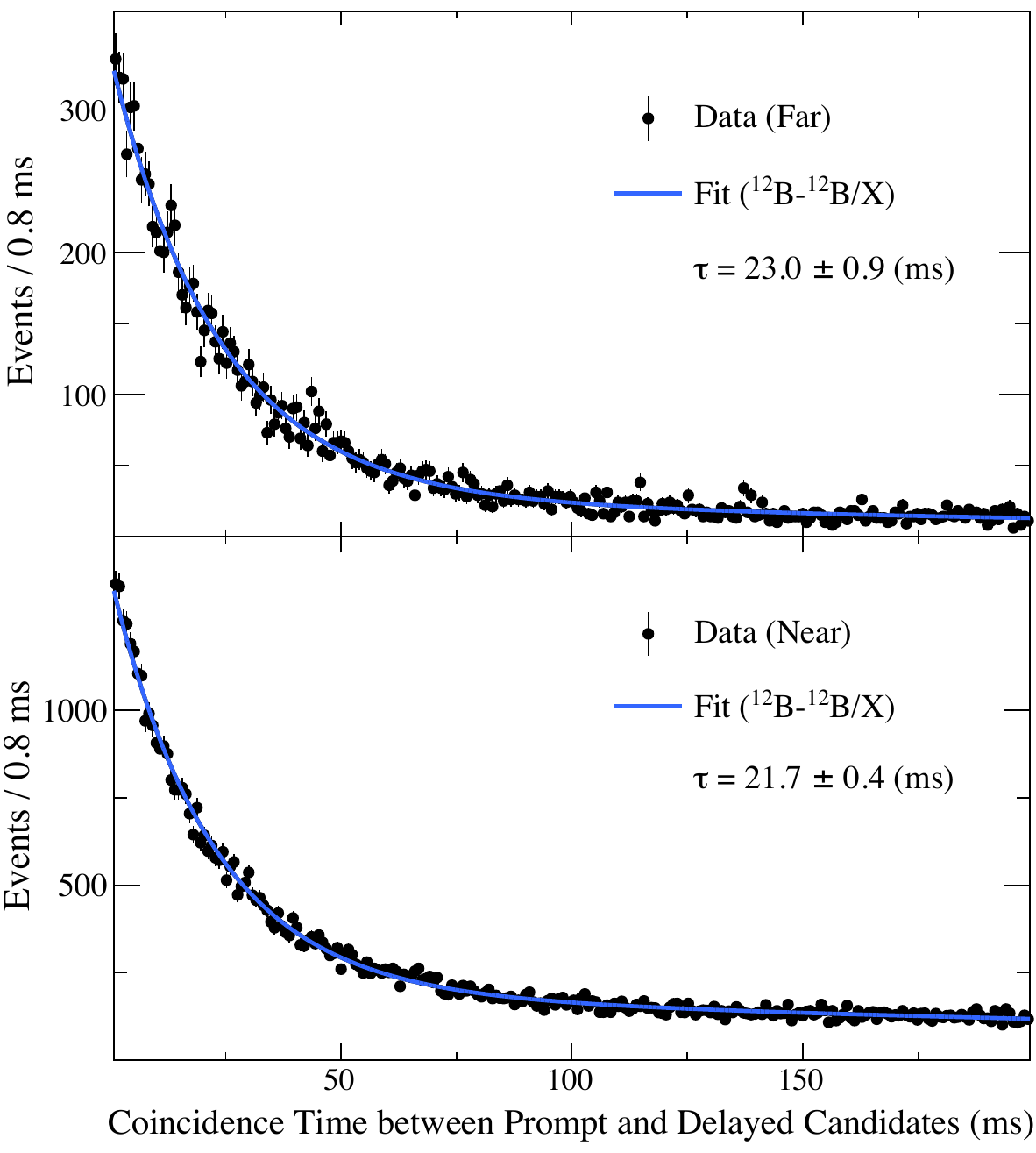} \caption{Time correlation between multiple $\rm ^{12}B$ decays produced by a single cosmic muon.}
\label{figure10}
\end{figure}

In order to obtain a $\rm ^{12}B$-$\rm ^{12}B/X$ enriched sample, the time coincidence between prompt and delayed candidates is expanded from 100\,$\mu \rm s$ to 200\,ms. Figure\,\ref{figure10} shows a time correlation of 23.0$\pm$0.9\,(21.7$\pm$0.4)\,ms between them at FD\,(ND). This indicates multiple $\rm ^{12}B$ production by a single cosmic muon. The longer time correlation component comes from either $\rm ^{12}B$ production by multiple cosmic muons or cosmogenic production of $\rm ^{12}B$ plus other longer-lifetime radioisotope. The multiple  $\rm ^{12}B$-$\rm ^{12}B$/X rate is measured from the time correlation distribution of Fig.\,\ref{figure10}. The rate of  $\rm ^{12}B$-$\rm ^{12}B$/X with their $\beta$-decay energies above 3\,MeV is measured to be $39.2\pm0.8$\,($9.1\pm0.3$)\,per day in the ND\,(FD).

\begin{figure}[h]
\includegraphics[width=0.48\textwidth]{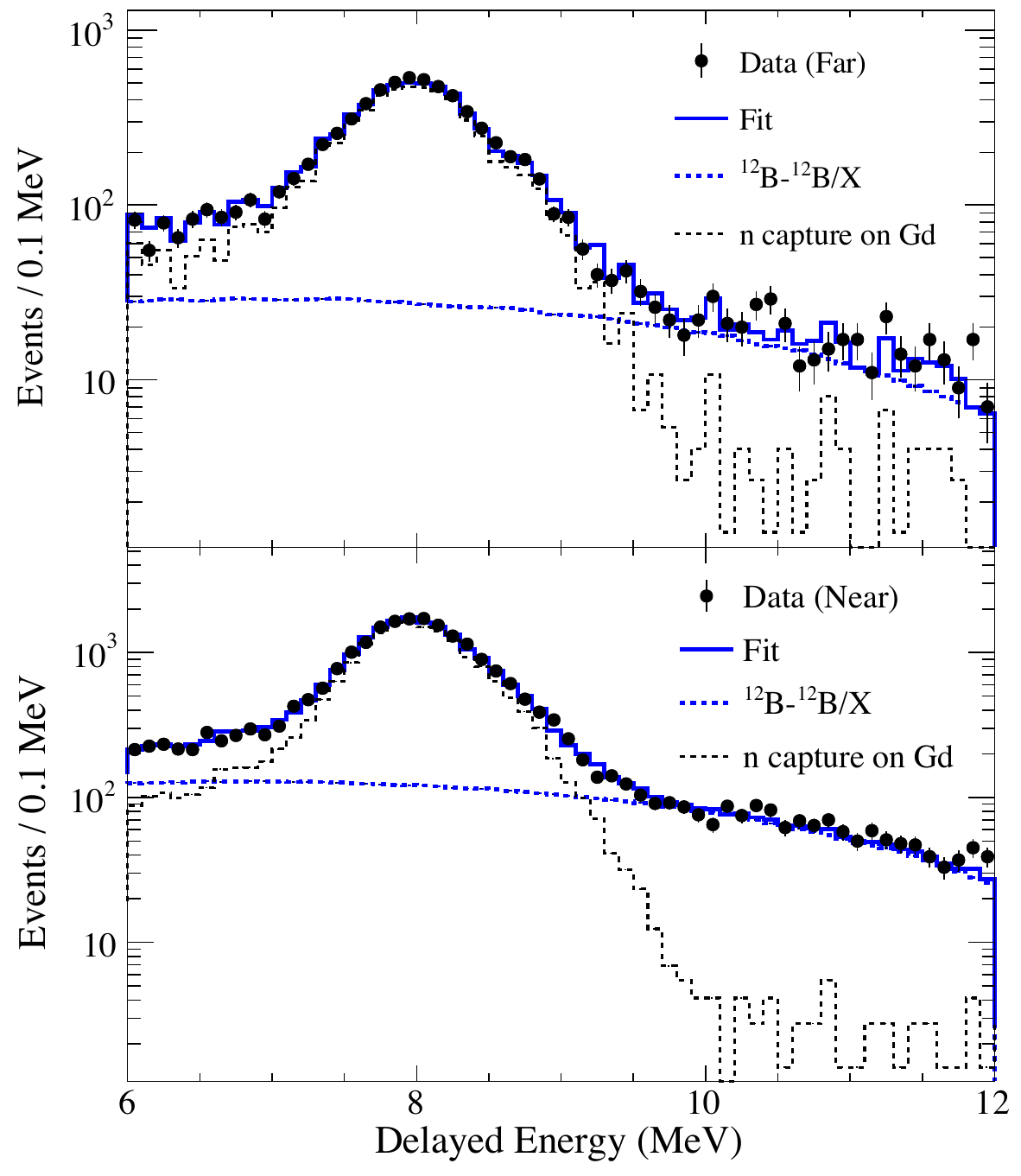} 
\caption{Delayed energy distribution of  $\rm ^{12}B$-$\rm ^{12}B$/X enriched $\beta$-n sample. The time coincidence in the sample is less than 3\,ms between prompt and delayed candidates.}
\label{figure11}
\end{figure}

   Figure\,\ref{figure11} shows the delayed energy distributions of the $\rm ^{12}B$-$\rm ^{12}B$/X enriched $\beta$-n sample where the time coincidence between prompt and delayed candidates is less than 3\,ms. The IBD delayed candidates show up as a distribution peaking at 8\,MeV, indicating $\gamma$ rays  from neutron capture on Gd. Their mean time coincidence is roughly 26\,$\mu$s. The delayed candidates from $\rm ^{12}B$ $\beta$ decays are clearly seen above 9\,MeV. The observed rate of $\rm ^{12}B$-$\rm ^{12}B$/X is obtained by a fit to the spectrum using measured spectral shapes of IBD delayed events and $\rm ^{12}B$/X $\beta$ decays. The $\rm ^{12}B$-$\rm ^{12}B$/X background rate remaining in the $\beta$-n candidate sample is obtained  from a delayed energy spectrum. The estimated rate at $8<E_{\rm p}<15$\,MeV is $0.05\pm0.01$\,per day at ND and $0.01\pm0.01$\,per day at FD. 
   
%---------------------------------------------
\subsection { Measurement of \Li+\HeS $\beta$-decay rate}
\begin{figure}[h]
\includegraphics[width=0.485\textwidth]{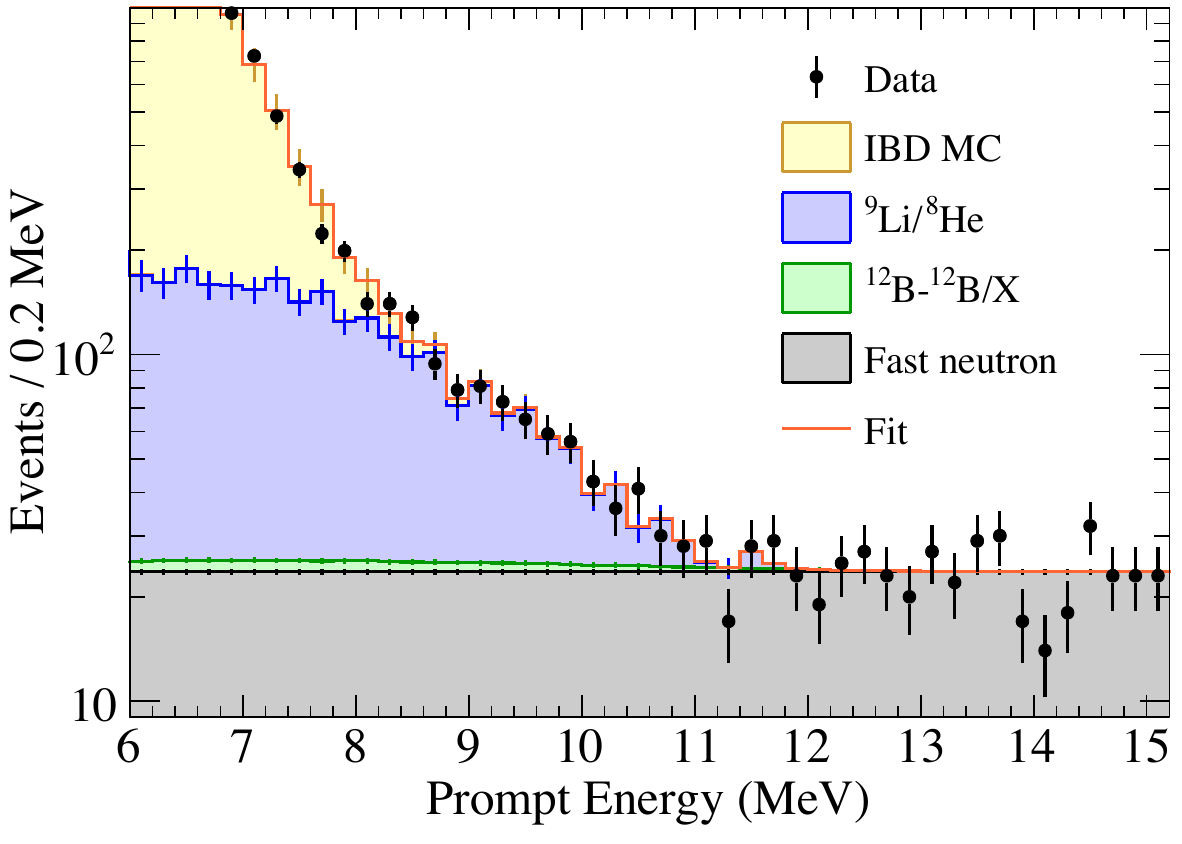}% bkg construction
\caption{Determination of \Li+\HeS $\beta$-decay rate at $8 < E_{\rm p}<15$\,MeV in ND. A spectral fit is performed by using the measured \Li+\HeS $\beta$-decay spectrum together with estimated background rates and spectra. The rate at $1.2 <E_{\rm p}< 8$\,MeV is estimated by extrapolating from the fit result using the measured \Li+\HeS $\beta$-decay spectrum.}
\label{figure12}
\end{figure}

The observed rate of \Li+\HeS $\beta$-decays at  $8< E_{\rm p} <15$\,MeV is determined by a spectral fit to the final $\beta$-n sample as shown in Fig.\,\ref{figure12}. The fit is performed by using the measured \Li+\HeS $\beta$-decays spectral shape together with estimated background rates and spectra. The obtained \Li+\HeS rate at $8 < E_{\rm p} <15\,$MeV is  1.46$\pm$0.07\,per day at ND and 0.37$\pm$0.04\,per day at FD as shown in  Table\,\ref{t:bkg}. The systematic error of the fit result comes from the uncertainties of the \Li+\HeS $\beta$-decay spectral shape, the IBD expectation, and the rest background contribution. The largest systematic error is due to the uncertainty of the measured \Li+\HeS $\beta$-decay spectrum at ND and the background at FD, and can be reduced with more data. The estimated background uncertainties  can also be reduced with more data.

\begin{table}[h!]
\begin{center}
\caption{\label{t:rate_err} Fractional errors of \LiS plus \HeS $\beta$-decay rates measured at $1.2 < E_{\rm p} < 15$\,MeV.}
\begin{tabular*}{0.48\textwidth}{@{\extracolsep{\fill}} lcc}
\hline
\hline
\multirow{2}{*}{Uncertainty source}& ND & FD  \\
& \multicolumn{2}{c}{\,($\%$)}   \\
\hline
%\hline
Statistics  at  8--15\,MeV  &2.12 & 4.32  \\
\LiS and \HeS spectrum at 1.2--15\,MeV & 2.26 & 2.26  \\
IBD at 8--15\,MeV   & 1.59  &  1.01  \\
Other backgrounds at 8--15\,MeV & 1.52  &  7.90  \\
\hline
Total error & 3.80 &  9.34 \\
\hline
\hline
\end{tabular*}
\end{center}
\end{table}

The rate at  $1.2 < E_{\rm p}< 8$\,MeV is estimated by extrapolating from the fit result using the measured \Li+\HeS $\beta$-decay spectrum.  The total \Li+\HeS $\beta$-decay rate for $1.2 < E_{\rm p} < 15$\,MeV is $11.92 \pm 0.25\,(\rm stat) \pm 0.38\,(\rm syst)$\,per day at ND and $3.04$$\pm$$ 0.13\,(\rm stat)$$ \pm $$0.25\,(\rm syst)$\,per day at FD. The fractional errors of the observed rates are shown in Table\,\ref{t:rate_err}. The largest error comes from the \LiS and \HeS spectrum uncertainty at ND and the background uncertainty at FD.

%---------------------------------------------
\vspace{0.4mm}
\vspace{0.4mm}
\vspace{0.4mm}

\subsection {Observed \LiS fraction}

\begin{figure}[h]
\begin{center}
\includegraphics[width=0.48\textwidth]{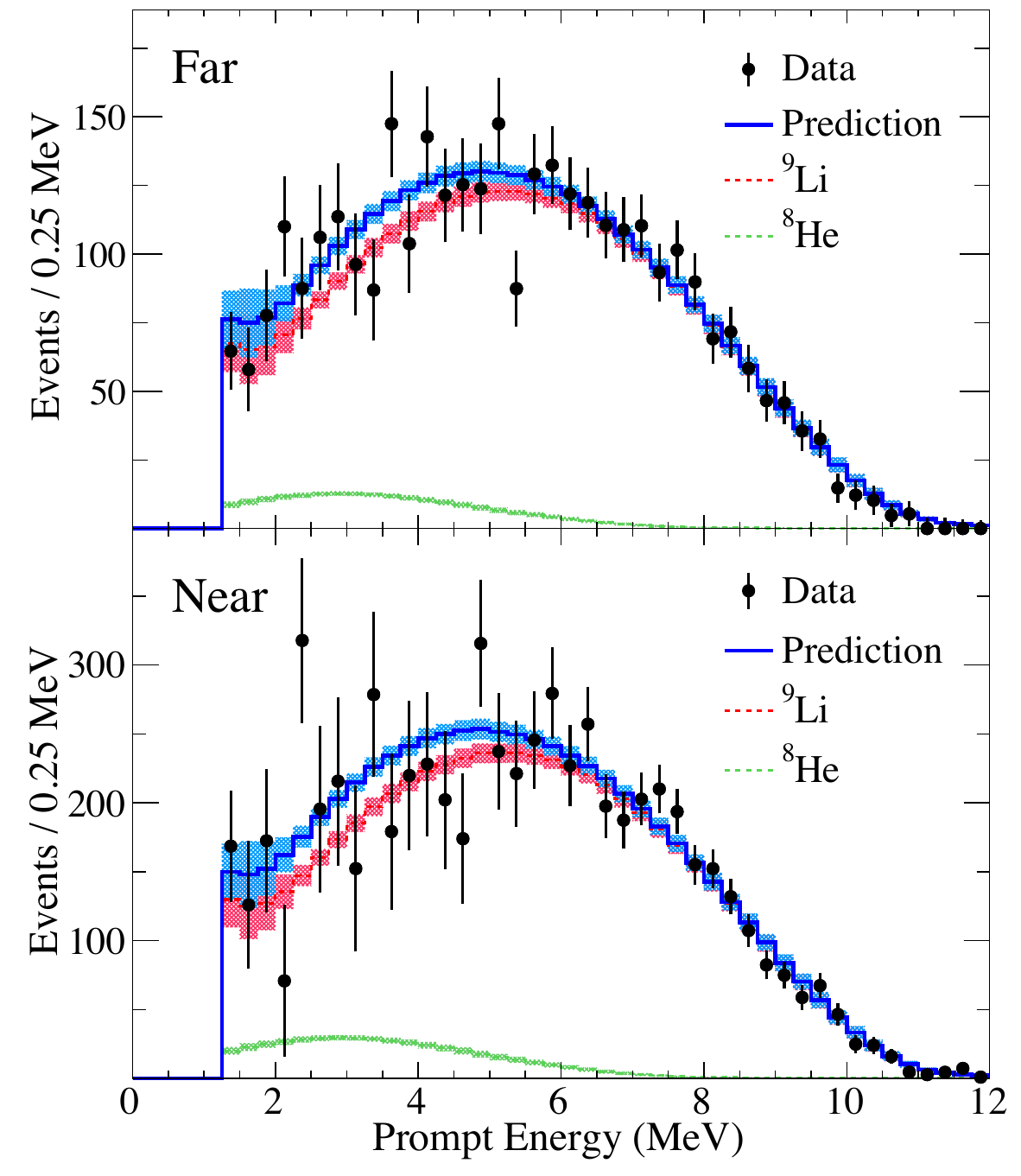} %fitting of fractional ratio of LiHe
\caption{\label{figure13} Fraction of  \LiS $\beta$ decays in the measured \Li+\HeS spectrum. A spectral fit to the measured spectrum is performed to obtain the fraction using the MC predicted $\beta$-decay spectra of \LiS and \He.}
\end{center}
\end{figure}

The $\beta$-n emitters are predominantly produced by \LiS over \He. Using the MC predicted $\beta$-decay spectra of \LiS and \He, the \LiS  contribution is determined by a fit to the measured \Li+\HeS $\beta$-decay spectrum, as shown in Fig.\,\ref{figure13}.

The obtained  fraction of  \Li\,(\He) $\beta$-n decay  is $(93.3$$\pm$$4.6)\%\,[(6.7$$\pm$$4.6)\%]$ for ND and $(94.2$$\pm$$4.1)\%$ $[(5.8$$\pm$$4.1)\%]$  for FD. The \HeS fractions of both detectors   are consistent with null rates indicating negligible contribution of \HeS to the observed $\beta$-n emitters. The errors come mostly from the statistical uncertainty and partly from the uncertainties associated with the predicted \LiS and \HeS $\beta$-decay spectra. Table\,\ref{t:fraction_err} shows the breakdown errors of the measured \LiS fraction. 

\begin{table}[h]
\begin{center}
\caption{\label{t:fraction_err} 
Errors of measured \LiS fraction. }
\begin{tabular*}{0.48\textwidth}{@{\extracolsep{\fill}} lcc}
\hline
\hline
Uncertainty source & ND\,(\%) & FD\,(\%)  \\

\hline
Data statistics    			& 4.0 & 3.2  \\
Branching ratio  			& 1.5 & 1.7  \\
Non-$\rm \beta$ particle    & 1.7 & 1.7  \\
Energy scale   				& 0.5 & 0.5  \\
Energy resolution    		& 0.2 & 0.3  \\
\hline
Total uncertainty   		& 4.6 & 4.1 \\
\hline
\hline
\end{tabular*}
\end{center}
\end{table}

%---------------------------------------------
\subsection {Cosmogenic \LiS and \HeS yields }

  The cosmogenic yields of \LiS or \HeS are determined from the observed $\beta$\--$\rm n$ rate from their decays. The \LiS yield per muon and mass in target region,  $Y\rm(^{9}Li)$, is given by the observed rate of \LiS and \HeS $\beta$-n decays$\,(R_{\rm \beta\textrm{\--}n}^{\rm obs})$, the measured fraction of \LiS decay$\,[f ({\rm ^{9}Li})]$, the detection efficiency of \LiS decay$\,[\epsilon ({\rm^{9}Li})]$, the $\beta$\--$\rm n$  branching ratio$\,(Br_{\beta\textrm{\--}n})$, the muon rate$\,(R_{\mu})$ passing through the target, the average muon track length in  target\,($\overline{L}_{\rm \mu}$) and the target density$\,(\rho)$\,:

   \vspace{1.5mm}

\begin{equation}
\begin{aligned}
Y\rm{(^{9}Li)}
= 
\textit{R}_{\rm{\beta \textrm{\--}n}}^{\rm{obs}}
\cdot
\frac
	{ 
	\textit{f}\,(\rm{^{9}Li})
	}{  
	\epsilon(\rm^{9}Li)
\cdot
	 \textit{Br}_{\rm \beta \textrm{\--} n}
\cdot
	\textit{R}_{\rm \mu}
\cdot
	\overline{\textit{L}}_{\rm \mu}
\cdot
	 \rho}_{\;\large{.}}
\end{aligned}
\end{equation}

   \vspace{1.8mm}

  The detection efficiencies of \LiS or \HeS $\beta$\--$\rm n$  decays are  obtained by using control samples and MC simulations. Most of the $\beta$\--$\rm n$  selection efficiencies are identical to those of the IBD event selection\,\cite{RENO:2016ujo} because of their common criteria for the prompt and delayed  candidates and coincidence between them. The only timing veto criteria against cosmic shower muons are avoided to collect maximum \LiS and  \HeS $\beta$\--$\rm n$  decays. The efficiency of prompt energy $E_{\rm p} >$ 1.2\,MeV requirement differs between \LiS and \HeS $\beta$ decays, and $(92.0$$\pm$$2.3)\%$ for \LiS and ($86.7$$\pm$$2.5)\%$ for \He. The errors come from the uncertainties of their expected $\beta$-decay spectra. The overall detection efficiency of \Li\,(\He) is ($49.6$$\pm$$1.5)\%\,((46.8$$\pm$$1.5)\%)$ for ND and $(56.2$$\pm$$1.6)\%\,((53.0$$\pm$$1.7)\%)$ for FD. The individual detection efficiencies are presented in Table\,\ref{t:det_eff}. 
   \vspace{1mm}

\begin{table}[h]
\caption{Detection efficiencies of \LiS or \HeS $\beta$-n decays.}
\label{t:det_eff}
\begin{center}
\begin{tabular*}{0.48\textwidth}{@{\extracolsep{\fill}} l c c c }
\hline\hline
\multirow{2}{*}{Selection requirement}&& \multicolumn{2}{c}{Efficiency\,(\%) }\\
&& ND  &   FD \\
\hline
$\rm Q_{{max}}/Q_{{tot}}$ &-& \multicolumn{2}{c}{100.00$\pm$0.02}  \\
Gd capture fraction &-& \multicolumn{2}{c}{85.0$\pm$0.8} \\
Spill-in &-& \multicolumn{2}{c}{101.3$\pm$0.9} \\
Time coincidence &-& \multicolumn{2}{c}{96.6$\pm$0.5 }\\
Spatial correlation &-& \multicolumn{2}{c}{ 100.00$\pm$0.03} \\
Delayed energy &-& \multicolumn{2}{c}{92.1$\pm$0.7} \\
\multirow{2}{*}{Prompt energy} & \multirow{2}{*}{\footnotesize{$\left\{\begin{array}{ll}{}^{9}\text{Li}\\{}^{8}\text{He}\end{array}\right.$}} &\multicolumn{2}{c}{92.0$\pm$2.3 }  \\                               &&\multicolumn{2}{c}{86.7$\pm$2.5 }  \\
\hline
Muon veto &-& 88.816$\pm$0.001 & 98.667$\pm$0.001 \\
Multiplicity &-& 99.991$\pm$0.001 & 99.155$\pm$0.006 \\
Trigger veto &-& 82.31$\pm$0.02 & 90.049$\pm$0.002 \\
${}^{252}\text{Cf}$ removal &-& 96.488$\pm$0.004 & 90.977$\pm$0.04 \\
Flasher removal &-& 99.93$\pm$0.01 & 99.51$\pm$0.03 \\
\hline
\multirow{2}{*}{Total} 
&\multirow{2}{*}{\footnotesize{$\left\{\begin{array}{ll}{}^{9}\text{Li}\\{}^{8}\text{He}\end{array}\right.$}}&49.6$\pm$1.5   & 56.2$\pm$1.6  \\
&&46.8$\pm$1.5   &53.0$\pm$1.7   \\
\hline
\hline
\end{tabular*}
\end{center}
\end{table}

 A detailed description of their estimation is given in Ref.\,\cite{RENO:2016ujo}.  The non $\beta$-n branching ratio of $\textit{Br}_{\rm \beta -n}$  needs to be taken into account to estimate the total production rate of \LiS and \HeS isotopes and is ($49.2$$\pm$$0.9$)\% for \LiS and ($84$$\pm$$1$)\% for \He\,\cite{Tilley:2004zz}. As described earlier, the rate of cosmic muons passing through the target region, $\textit{R}_{\rm \mu}$, is estimated to be $61.8$$\pm$$0.7$\,$\rm s^{-1}$ at ND and $6.9$$\pm$$0.1$\,$\rm s^{-1}$ at FD. An average muon track length in target\,($\overline{\textit{L}}_{\rm \mu}$) is obtained as $201.1$$\pm$$3.9$\,cm at ND and $197.4$$\pm$$3.9$\,cm at FD. Table\,\ref{t:yield} shows values necessary for converting the observed rate into a yield.

\begin{table}[h]
\begin{center}
\caption{\label{t:yield} List of values used for converting observed rate into yield.}
\begin{tabular*}{0.48\textwidth}{@{\extracolsep{\fill}} lccc}
\hline
\hline
  		&	& ND 			& FD  \\
\hline
$R_{\rm \beta-n}\;\,(\rm /day)$  &-& 11.92$\pm$0.45 &  3.04$\pm$0.28 \\
$R_{\mu} \;\,(\rm \rm /sec)$ &-& 61.84$\pm$0.71 &  6.94$\pm$0.08 \\
$\overline{\textit{L}}_{\rm \mu} \;(\rm cm)$&-& 201.1$\pm$3.9 &  197.4$\pm$3.9 \\
$\rho\;\,(\rm g/cm^3)$  &-& \multicolumn{2}{c}{0.856$\pm$0.001} \\
\multirow{2}{*}{$\epsilon \;\,(\%)$} 
&\multirow{2}{*}{\footnotesize{$\left\{\begin{array}{ll}{}^{9}\text{Li}\\{}^{8}\text{He}\end{array}\right.$}}& 49.6$\pm$1.5 &  56.2$\pm$1.6  \\
&& 46.8$\pm$1.5 &  53.0$\pm$1.7  \\
\multirow{2}{*}{$f \;\,(\%)$ }    
&\multirow{2}{*}{\footnotesize{$\left\{\begin{array}{ll}{}^{9}\text{Li}\\{}^{8}\text{He}\end{array}\right.$}}& 93.3$\pm$4.6 &  94.2$\pm$4.1 \\
&& 6.7$\pm$4.6 &  5.8$\pm$4.1 \\

\multirow{2}{*}{$ Br_{\rm \beta-n} \;\,(\%)$} 
&\multirow{2}{*}{\footnotesize{$\left\{\begin{array}{ll}{}^{9}\text{Li}\\{}^{8}\text{He}\end{array}\right.$}}& \multicolumn{2}{c}{50.8$\pm$0.9} \\ 
&& \multicolumn{2}{c}{16$\pm$1} \\ 
\hline
\hline
\end{tabular*}
\end{center}
\end{table}

   The cosmogenic isotope yields are obtained in a unit of $\rm 10^{-8}$ $\mu^{-1}g^{-1}cm^{2}$. The \LiS yield is measured to be $4.80$$\pm$$0.36$ for ND and $9.9$$\pm$$1.1$ for FD. The \HeS yield is measured to be $1.15$$\pm$$0.81$ for ND and $2.1$$\pm$$1.5$ for FD, and rather consistent with a null signal within its large error. The total yield of \Li+\HeS is obtained as $5.95$$\pm$$0.65$ for ND and $12.0$$\pm$$1.6$ for FD. Note that the \Li+\HeS yield with a subsequent $\beta$-n decay is insensitive to the individual isotope fraction of a large uncertainty, and obtained as $2.62$$\pm$$0.14$ for ND and $5.36$$\pm$$0.54$ for FD. The measured yields of \Li, \HeS and \Li+\HeS are listed in Table\,\ref{t:cosmogenic_yield}.
   \vspace{1.2mm}

 The cosmogenic \Li\,(\He) production rate is estimated \\from the yield, and obtained as $2.77$$\pm$$0.20$$\textcolor{red}{\,(}0.66$$\pm$$0.47)$ $\rm ton^{-1} day^{-1}$ for ND and $0.628$$\pm$$0.068$\,$(0.130$$\pm$$0.092)$ $\rm ton^{-1}$ $\rm day^{-1}$ for FD.  The cross section of \Li$\,$(\He) production is derived from the yield and the target mass of $\rm ^{12}C$, and obtained as $0.96$$\pm$$0.07$\,$(0.23$$\pm$$0.16)$\,$\rm \upmu b$ for ND and $1.97$$\pm$$0.22$\,$(0.41$$\pm$$0.29)$\,$\rm \upmu b$ for FD. The cosmogenic yields, production rates, and cross sections for both detectors are given in Table$\,$\ref{t:comp_other_exp}  where other experimental results are compared.

\begin{table}[h]
\caption{Measured cosmogenic yields of \Li, \HeS and \Li+\He. The first error is statistical and the second one is systematical.}
\label{t:cosmogenic_yield}
\begin{center}
\begin{tabular*}{0.48\textwidth}{@{\extracolsep{\fill}} l c c }
\hline
\hline
\multirow{2}{*}{Isotope}&\multicolumn{2}{c}{Yield\,$\rm (\times  10^{-8} \mu^{-1}g^{-1}cm^{2})$ }  \\
&ND &FD\\
\hline
$^{9}\text{Li}$&$\rm4.80$$\pm$$0.23$$\pm$$0.28$&$\rm9.89$$\pm$$0.54$$\pm$$0.95$\\
$^{8}\text{He}$&$\rm1.15$$\pm$$0.69$$\pm$$0.42$&$\rm 2.1$$\pm$$1.1$$\pm$$0.9$\\
$^{9}\text{Li}$+$^{8}\text{He}$&$\rm5.95$$\pm$$0.50$$\pm$$0.41$&$\rm12.0$$\pm$$0.94$$\pm$$1.26$\\
$^{9}\text{Li}$+$^{8}\text{He}\rightarrow \beta+n$&$\rm2.62$$\pm$$0.06$$\pm$$0.12$&$\rm5.36$$\pm$$0.23$$\pm$$0.48$\\
\hline
\hline
\end{tabular*}
\end{center}
\end{table}

\begin{table*}
\begin{ruledtabular}
\begin{tabular}{l cc cc cc cc}
 \multicolumn{2}{l}{\multirow{2}{*}{Experiment\,\;\;\;\;\;\;\;\;\;Detector}}& $\overline{\textit{E}}_{\rm \mu}$ 
&  $ Y_{\rm Li}$  &  $ Y_{\rm He}$  &  $ R_{\rm Li}^{\rm prod}$  &  $ R_{\rm He}^{\rm prod}$  &  $ \sigma_{\rm Li}$  &  $ \sigma_{\rm He}$\\
&  &\,($\rm GeV$)
& \multicolumn{2}{c}{\,($\rm \times 10^{-8}  \mu^{-1}g^{-1}cm^{2}$)} 
& \multicolumn{2}{c}{\,($\rm ton^{-1} day^{-1}$)} 
& \multicolumn{2}{c}{\,($\rm \upmu b$)}  \\
\hline
 \multicolumn{2}{l}{ \multirow{2}{*}{RENO}\;\;\;\;\;\;\;\;\;\;\;\;\,\;\;\;\; \multirow{2}{*}{\scriptsize{$\left\{\begin{array}{ll}\small{\text{ND}}\\ \small{\text{FD}} \end{array}\right.$}} }
  & 33.1$\pm$2.3    & 4.80$\pm$0.36 & 1.15$\pm$0.81  & 2.77$\pm$0.20 & 0.66$\pm$0.47 & 0.96$\pm$0.07 & 0.23$\pm$0.16 \\
					 
  &   & 73.6$\pm$4.4   & 9.89$\pm$1.09 &2.05$\pm$1.45  & 0.628$\pm$0.068 &0.130$\pm$0.092 & 1.97$\pm$0.22 & 0.41$\pm$0.29 \\
				   	           
% \multicolumn{2}{l}{ \multirow{3}{*}{Daya Bay}\;\;\;\;\;\;\;\;\,\;\;\; \multirow{3}{*}{\tiny{$\left\{\begin{array}{ll}\small{\text{EH01}}\\ \small{\text{EH02}} \\ \small{\text{EH03}} \end{array}\right.$}} }
 %   & 63.9$\pm$3.8 &  7.66$\pm$0.81  & - & - & - & - & -\\
%				           &  & 64.7$\pm$3.9 &  7.72$\pm$0.91 &- & - & - & - & -\\
%				           &  & 143.0$\pm$8.6 &  15.65$\pm$1.85 &- & - & - & - & -\\
				           
 \multicolumn{2}{l}{ \multirow{2}{*}{Double Chooz}\;\;\;\;\, \multirow{2}{*}{\scriptsize{$\left\{\begin{array}{ll}\small{\text{ND}}\\ \small{\text{FD}} \end{array}\right.$}} }
  & 32.1$\pm$2.0 &5.51$\pm$0.52 
&$<~$4.96 & 1.73$\pm$0.16 & $<~$1.56 & 1.10$\pm$0.10 & $<~$0.99 \\
				   	         & &63.7$\pm$5.5 &7.90$\pm$0.51 
&0.77$\pm$1.61& 0.48$\pm$0.03 & 0.05$\pm$0.10 & 1.57$\pm$0.10 & 0.15$\pm$0.32 \\
KamLAND & - &  260$\pm$8 &  22$\pm$2    &7$\pm$4 & 0.0028$\pm$0.0002 & 0.001$\pm$0.00005 & - & - \\
Borexino & -  & 283$\pm$19 &  29$\pm$3  &$<~$2.0 & 0.00083$\pm$0.00009 & $<~$0.00042 & - & - \\

\end{tabular}
\end{ruledtabular}
\caption{\label{t:comp_other_exp}  Yields, production rates, and cross sections of cosmogenic \LiS and \HeS radioisotopes. The measured values are also given for 
Borexino\,\cite{Borexino:2013cke}, 
%Daya Bay\,\cite{DayaBay:2016ggj}, 
Double Chooz\,\cite{DoubleChooz:2018kvj} and KamLAND\,\cite{KamLAND:2009zwo}. Daya Bay's observed results\,\cite{DayaBay:2016ggj} are not included because of missing absolute detection efficiency and acceptance while Ref.\,\cite{DoubleChooz:2018kvj} presents estimated values assuming full contribution of \Li.}
\end{table*}

\begin{figure}
\includegraphics[width=0.485\textwidth]{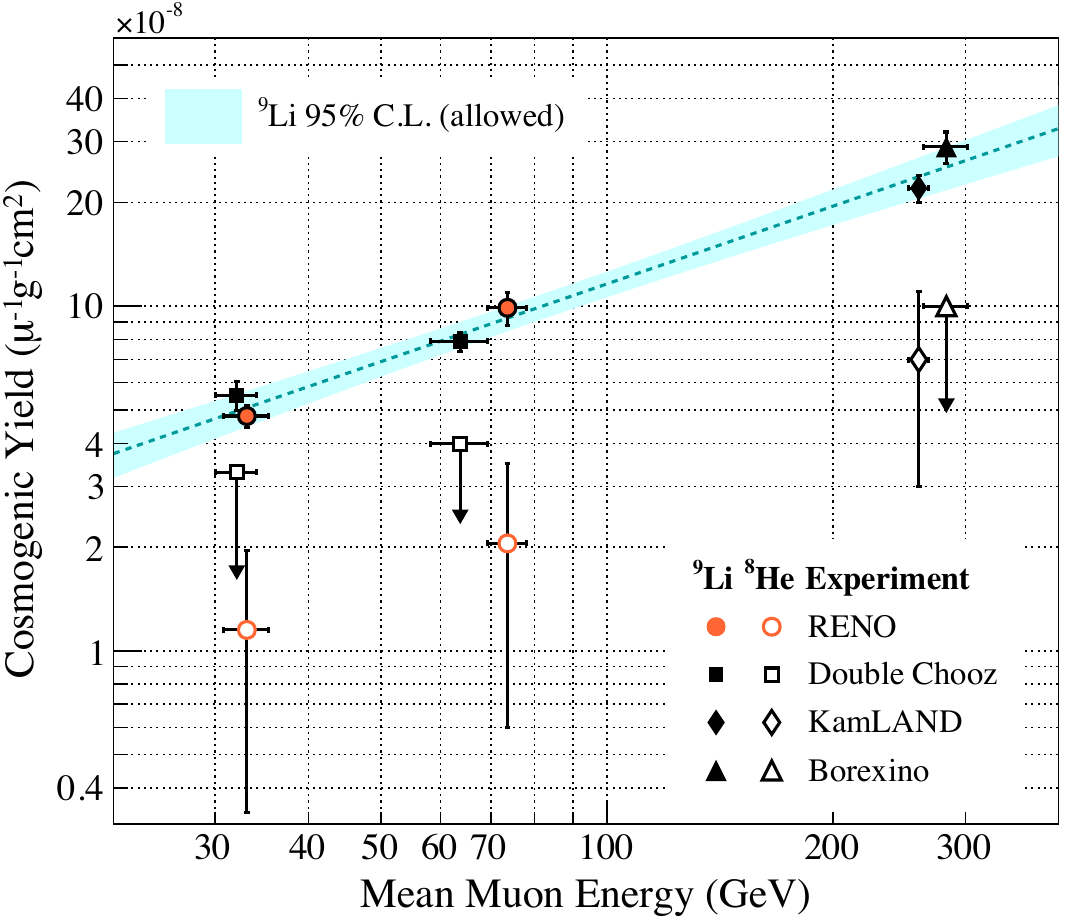}
\caption{Measured yields of cosmogenic \LiS and \HeS at RENO’s ND and FD. They are compared with those of Double Chooz, Borexino and KamLAND. The arrows present an upper limit of 2\,$\sigma$. A fit to the \LiS yields obtains the power-law exponent of  $\alpha = 0.75$$\pm$$0.05$ with $\rm \chi^2/NDF=0.98$. }
\label{figure14}
\end{figure}

 The cosmogenic yields at underground sites are expected to have a dependence on the mean muon energy\,($\overline{\textit{E}}_{\rm \mu}$),

\begin{equation}
\begin{aligned}
Y
= 
Y_{\rm 0}
\left(
			\frac
				{ 
				\overline{\textit{E}}_{\rm \mu}
				}{  
				\rm 1\,GeV
				}
\right)^{\alpha},
\end{aligned}
\end{equation}

\noindent where $\alpha$ is a power-law exponent as a function of $\overline{\textit{E}}_{\rm \mu}$\,\cite{Hagner:2000xb}. As described in Section III.A, the mean muon energies are $33.1$$\pm$$2.3$ and $73.6$$\pm$$4.4$\,GeV for the ND and FD sites, respectively, obtained from a MUSIC simulation. Figure\,\ref{figure14} shows comparison of the measured \LiS or \HeS yields with other measurements from %Daya Bay\,\cite{DayaBay:2016ggj}, 
Borexino\,\cite{Borexino:2013cke}, Double Chooz\,\cite{DoubleChooz:2018kvj} and KamLAND\,\cite{KamLAND:2009zwo}. The yields of Double Chooz are obtained from the probability of neutron production within 1\,ms and a certain distance with respect to a preceding muon\,\cite{DoubleChooz:2018kvj}. 
   \vspace{0.5mm}

   \vspace{0.5mm}

%The \textcolor{red}{$\beta$-n rate} of Daya Bay are obtained by summing \textcolor{red}{an observed $\beta$-n rate} due to shower muons and an estimated rate due to non-showering muons, assuming \LiS production only. The shower muon induced rate is measured from an elapsed time distribution from a preceding muon and the non-shower induced rate is estimated from neutron capture rate within 200\,$\mu s$  from a preceding muon\,\cite{DayaBay:2016ggj}. 

These earlier measurements are based on time and spatial correlation with a preceding cosmic muon. However, the association becomes uncertain when it comes to a high muon rate relative to the \LiS or \HeS lifetimes at a shallow underground. The RENO  measurement does not rely on the time and spatial correlation with a preceding muon, but is obtained from direct counting the \LiS and  \HeS $\beta$-n emitters at  $8 < E_{\rm p} <15$\,MeV by a fit using a measured spectrum. Therefore, this result provides a direct measurement of the cosmogenic  \LiS and  \HeS yields at shallow overburdens to obtain an accurate value of the power-law exponent. A fit to the cosmogenic \LiS yields at several underground locations finds  $\alpha = 0.75$$\pm$$0.05$ and $Y_{0} \rm = (0.37$$\pm$$0.08)\times 10^{-8}$ $\rm \mu^{-1} g^{-1} cm^{2}$ with $\rm \chi^2/NDF=0.98$ as shown in Fig.\,\ref{figure14}.

%----------------------- Summary
\section{Summary}
   
The $\beta$-n emitters from \LiS and \HeS radioisotopes are produced as spallation products of cosmic muons in the RENO ND and FD. Their observed rates are obtained by a spectral fit without their muon time and spatial information. A relative \textcolor{black}{fraction of the two cosmogenic isotopes} is measured by MC expected spectra of their $\beta$ decays, and thus allows a direct measurement of the individual \LiS and \HeS yields. Other cosmogenic radioisotopes such as \BS and \NS are also \textcolor{black}{observed,} and their contribution to the $\beta$-n sample is estimated. 
    \vspace{0.5mm}

The cosmogenic \LiS yields at the two underground detector sites show a clear relationship with the mean muon energy. A well-behaved power-law dependence of the \LiS yield is obtained as a function of the mean muon energy from combining available other measured values. The power-law relationship provides a useful prediction of \LiS and \HeS $\beta$-n emitting background rates for $\rm ^{12}C$ based underground detectors.
 
%----------------------- Acknowledgements
\section*{Acknowledgements}
The RENO experiment is supported by the National Research Foundation of Korea (NRF) Grants No. 2009-0083526, No. 2019R1A2C3004955, 2021R1A2C1013661, and 2022R1A3B1078756 funded by the Korean Ministry of Science and ICT. Some of us have been supported by a fund from the BK21 of NRF. This work was partially supported by the New Faculty Startup Fund from Seoul National University. We gratefully acknowledge the cooperation of the Hanbit Nuclear Power Site and the Korea Hydro \& Nuclear Power Co., Ltd. (KHNP). We thank KISTI for providing computing and network resources through GSDC, and all the technical and administrative people who greatly helped in making this experiment possible. \\

%----------------------- reference
\input{RENO_LiHe.bbl}
\end{document}

%% file: RENO_LiHe.bbl
%apsrev4-2.bst 2019-01-14 (MD) hand-edited version of apsrev4-1.bst
%Control: key (0)
%Control: author (8) initials jnrlst
%Control: editor formatted (1) identically to author
%Control: production of article title (0) allowed
%Control: page (0) single
%Control: year (1) truncated
%Control: production of eprint (0) enabled
\providecommand{\noopsort}[1]{}\providecommand{\singleletter}[1]{#1}%